\documentstyle[preprint,eqsecnum,amsfonts,amsmath,aps]{revtex}

\allowdisplaybreaks

\begin{document}
\draft

\title{Third post-Newtonian dynamics of compact binaries:\\
Noetherian conserved quantities and equivalence between\\ 
the harmonic-coordinate and ADM-Hamiltonian formalisms}

\author{Vanessa C. de Andrade}
\address{D\'epartement d'Astrophysique Relativiste et de
Cosmologie, \\ 
Centre National de la Recherche Scientifique (UMR 8629),\\
Observatoire de Paris, 92195 Meudon Cedex, France}

\author{Luc Blanchet}
\address{D\'epartement d'Astrophysique Relativiste et de Cosmologie,\\
Centre National de la Recherche Scientifique (UMR 8629),\\
Observatoire de Paris, 92195 Meudon Cedex, France}

\author{Guillaume Faye}
\address{Theoretisch-Physikalisches Institut,
Friedrich-Schiller-Universit\"at,\\
Max-Wien-Pl. 1, 07743 Jena, Germany}

\date{\today}
\maketitle

\begin{abstract}
A Lagrangian from which derive the third post-Newtonian (3PN)
equations of motion of compact binaries (neglecting the radiation
reaction damping) is obtained. The 3PN equations of motion were
computed previously by Blanchet and Faye in harmonic coordinates. The
Lagrangian depends on the harmonic-coordinate positions, velocities
and accelerations of the two bodies. At the 3PN order, the appearance
of one undetermined physical parameter $\lambda$ reflects an
incompleteness of the point-mass regularization used when deriving the
equations of motion. In addition the Lagrangian involves two
unphysical (gauge-dependent) constants $r'_1$ and $r'_2$ parametrizing
some logarithmic terms. The expressions of the ten Noetherian
conserved quantities, associated with the invariance of the Lagrangian
under the Poincar\'e group, are computed. By performing an
infinitesimal ``contact'' transformation of the motion, we prove that
the 3PN harmonic-coordinate Lagrangian is physically equivalent to the
3PN Arnowitt-Deser-Misner Hamiltonian obtained recently by Damour,
Jaranowski and Sch\"afer.
\end{abstract}

\section{Motivation and relation to other works}

The long-standing problem of the gravitational dynamics of compact
bodies has become very important in recent years because of the need
to construct accurate templates for detecting the gravitational waves
from inspiralling compact binaries in future experiments like LIGO and
VIRGO
\cite{3mn,CFPS93,DIS98}. Concerning the two-body problem the current
state of the art is the 3PN approximation, corresponding to the
inclusion of all the relativistic corrections up to the order $1/c^6$
(where $c$ is the velocity of light) with respect to the Newtonian
acceleration. Up to the 2.5PN or $1/c^5$ approximation the equations
of motion are well known, as they have been derived by many different
methods with complete agreement on the result
\cite{LD17,EIH,F50,BeDD81,DD81a,DD81b,D82a,D82b,DS85,Kop85,GKop86,S85,S86,BFP98}. 
They have already been used for constructing the 2.5PN-accurate
templates of inspiralling compact binaries
\cite{BDI95,WW96,B96}.

To the 3PN order, the problem of equations of motion has been pursued
by two groups working independently with different methods: on one
hand, Jaranowski and Sch\"afer \cite{JS98,JS99} and Damour, Jaranowski
and Sch\"afer
\cite{DJS00a,DJS00b,DJS00c} employ the Arnowitt-Deser-Misner (ADM)
Hamiltonian formulation of general relativity; on the other hand,
Blanchet and Faye
\cite{BF00,BFreg,BFregM,BFeom} work iteratively with the Einstein
field equations in harmonic coordinates. Both groups use a
regularization based on Hadamard's concept of ``partie finie'' to
overcome the problem of the infinite self-field of point-like
particles. However the details are actually different; notably the
second group developed for this problem an extended version of the
Hadamard regularization and a theory of generalized functions
\cite{BFreg,BFregM}. Both groups found that there remains one and only
one physical constant, $\omega_{\rm static}$ in the ADM-Hamiltonian
formalism
\cite{JS98,JS99,DJS00a,DJS00b,DJS00c} and $\lambda$ in the
harmonic-coordinate approach
\cite{BF00,BFreg,BFregM,BFeom}, that is left
un-determined by the point-mass regularization. Furthermore, in the
harmonic-coordinate approach, the equations of motion (obtained in
Ref. \cite{BFeom}) depend on two additional constants $r'_1$ and
$r'_2$ parametrizing some logarithmic terms, but these constants are
not physical in the sense that they can be removed by a coordinate
transformation. The aim of the present paper is three-fold: (i) to
present the Lagrangian of the 3PN dynamics of the compact binary in
harmonic coordinates, (ii) to obtain explicitly from it the ten
Noetherian conserved integrals of the motion in harmonic coordinates,
(iii) to exhibit a contact transformation of the harmonic-coordinate
motion to some pseudo-ADM coordinates in order to compare our results
\cite{BF00,BFreg,BFregM,BFeom} with the ones obtained by the other
group
\cite{JS98,JS99,DJS00a,DJS00b,DJS00c}.

Concerning (i), we find a generalized Lagrangian (i.e. depending on
the positions, velocities and accelerations of the bodies) whose
variation yields the conservative part of the 3PN equations of motion
in harmonic coordinates as found in Ref. \cite{BFeom}. Our second
point (ii) is to use the fact that the Lagrangian incorporates the ten
symmetries of the Poincar\'e group (notably the boost symmetry) to
compute the ten integrals corresponding to the energy, the linear and
angular momenta, and the center-of-mass position. In particular, we
find that the energy agrees with the previous result of
Ref. \cite{BFeom}. As all these integrals will probably be needed in
future work we choose to display them explicitly, despite the length
of the expressions. We also give the balance equations they satisfy
when the radiation reaction effect is turned on. Finally, the result
of point (iii) is that there exists a unique contact transformation of
the harmonic-coordinate dynamical variables that changes the
generalized Lagrangian into an ordinary Lagrangian (depending on
positions and velocities) whose associated 3PN Hamiltonian matches
exactly the one given by Damour, Jaranowski and Sch\"afer
\cite{DJS00b}. This proves the complete equivalence of the results 
obtained from the two (rather different) methods followed by the two
groups, and constitutes a strong support of the validity of both
methods. This equivalence has also been shown independently by the
other group \cite{DJS00c} (who presents also the formulas needed for
computing the conserved quantities). Notice that it holds if and only
if the un-determined constant $\lambda$ in the harmonic-coordinate
formalism and the ambiguity constant $\omega_{\rm static}$ in the ADM
Hamiltonian are related to each other by

\begin{equation}\label{1}
\omega_{\rm static}=-\frac{11}{3}\lambda-\frac{1987}{840}\;,
\end{equation}
a result already obtained in Ref. \cite{BF00} on the basis of the
comparison of the invariant energy of binaries moving on circular
orbits. Likely the appearance of the unknown constant $\lambda$ is
not due to a real physical ambiguity, but is associated with an
incompleteness of the point-mass regularization. It is probably
related to the fact that, starting from the 3PN order, many separate
integrals constituting the equations of motion of extended bodies
would depend on the internal structure of the objects (e.g. their
density profile), even in the limiting case where the radius of
the objects tends to zero. Further work is needed to compute the
precise value of $\lambda$. On the other hand, the constants $r'_1$
and $r'_2$ occuring in the harmonic-coordinate Lagrangian disappear
from the ADM-Hamiltonian (where there are no logarithms), in
accordance with the fact that they are pure gauge.

The plan of this paper is as follows. In Section II, motivated by the
striking equivalence between the (regularization-related) unknown
constants $\lambda$ and $\omega_{\rm static}$, we discuss our method
of point-mass regularization and contrast it to the method advocated
in
\cite{JS98,JS99,DJS00a,DJS00b,DJS00c}. Section III is devoted 
to the theoretical investigations. First we recall the theory of
Noetherian conserved quantities in the case of a generalized
Lagrangian, and next we show how to eliminate the accelerations in the
harmonic-coordinate Lagrangian by a contact transformation at the 3PN
order. The reader interested only in the results at the 3PN order can
go directly to Section IV, where we present the closed-form
expressions of the Lagrangian and the conserved energy, momenta and
center of mass in harmonic coordinates, and give the result for the
contact transformation as well as the final expressions for the
Lagrangian and Hamiltonian in pseudo-ADM coordinates.

\section{Discussion on the point-mass regularization}

The equivalence between the respective formalisms of
\cite{JS98,JS99,DJS00a,DJS00b,DJS00c} and
\cite{BF00,BFreg,BFregM,BFeom} is interesting because the two groups
have adopted some different approaches regarding the point-mass
regularization (chosen in both cases to be based on the Hadamard
concept of ``partie finie'' of a singular function or a divergent
integral
\cite{Hadamard,Schwartz}). Essentially the group 
\cite{JS98,JS99,DJS00a,DJS00b,DJS00c} introduced 
systematically some ``ambiguity'' parameters in the ADM Hamiltonian
whenever the standard Hadamard regularization yielded inconsistent
results, while the group \cite{BF00,BFreg,BFregM,BFeom} looked for the
most general solution allowed by some basic physical requirements and
following from a new, mathematically consistent, Hadamard-type
regularization.

More precisely, in our approach \cite{BF00,BFreg,BFregM,BFeom}, we
adopted some specific variants of the Hadamard regularization which
were devised specifically for this problem \cite{BFreg,BFregM}. Let
$F$ be a function which is singular at two isolated points ${\bf y}_1$
and ${\bf y}_2$, and is smooth everywhere else; ${\bf y}_1$ and ${\bf
y}_2$ are the positions of the particles in harmonic coordinates at
some given instant $t$. The Hadamard partie finie of $F$ at the point
${\bf y}_1$, denoted $(F)_1$, is defined as the angular average over
all directions of approach to ${\bf y}_1$ of the finite term (zeroth
order) in the singular expansion of the function around this point. We
found that this definition yields a natural extension of the notion of
Dirac distribution at the location of a singular point, that we
constructed by means of the Riesz delta-function \cite{Riesz}. As a
result, the ``partie finie delta-function'' at the point 1, denoted
${\rm Pf}\delta_1$ where $\delta_1\equiv \delta({\bf x}-{\bf y}_1)$,
is the linear form defined on the set of singular functions of the
type $F$, that associates to any $F$ the real number $(F)_1$ (see
Eq. (6.9) in
\cite{BFreg}).  Using an integral notation this means that $\int
d^3{\bf x}~F.{\rm Pf}\delta_1=(F)_1$. (The partie finie delta-function
${\rm Pf}\delta_1$ constitutes a mathematically well-defined version
of the so-called ``good delta function'' of Infeld \cite{InfeldP}.) In
our derivation of the equations of motion at 3PN order, this
prescription is employed systematically to compute all the
``compact-support'' integrals, whose integrand is made of the product
of a singular potential with some mass density localized on the two
particle world-lines.

By applying the latter definition to the product $FG$ we obtain $\int
d^3{\bf x}~FG.{\rm Pf}\delta_1=(FG)_1$, which permits us to give a
sense to the more complicated object $F.{\rm Pf}\delta_1\equiv {\rm
Pf}(F\delta_1)$, composed of the product of a delta-pseudo-function
with a function which is singular on its support (such a product being
ill-defined in the standard distribution theory). Namely, ${\rm
Pf}(F\delta_1)$ is the linear form which associates to any function
$G$ the real number $(FG)_1$. It is important to realize that ${\rm
Pf}(F\delta_1)\not=(F)_1{\rm Pf}\delta_1$ in general. This is an
immediate consequence of the so-called ``non-distributivity'' of the
Hadamard partie finie, namely the fact that $(FG)_1\not=(F)_1(G)_1$
for two singular functions $F$ and $G$ in general. As an exemple taken
from \cite{BFP98}, we have $(U^4)_1=[(U)_1]^4+2[(U)_1]^2[(U)_2]^2$,
where $U=Gm_1/r_1+Gm_2/r_2$ denotes the Newtonian potential of two
particles (with $r_1=|{\bf x}-{\bf y}_1|$ and $r_2=|{\bf x}-{\bf
y}_2|$). In the post-Newtonian iteration one can check that the
functions involved become singular enough so that the
non-distributivity plays an actual role at the 3PN order: for
instance, in the example above, $U^4$ will appear in the metric
coefficient $g_{00}$ with a factor $1/c^8$ in front, which indeed
corresponds to the 3PN order. However, there is no problem linked with
the non-distributivity in the equations of motion up to the 2.5PN
approximation \cite{BFP98}. Therefore, from the 3PN order (but only
from that order), it is a mathematically inconsistent regularization
prescription to assume at once that $\int d^3{\bf x}~F.{\rm
Pf}\delta_1=(F)_1$ and ${\rm Pf}(F\delta_1)=(F)_1{\rm Pf}\delta_1$.
Faced with this problem, the authors
\cite{JS98,JS99,DJS00a,DJS00b,DJS00c} have advocated that the breakdown of
the distributivity of the Hadamard regularization at the 3PN order is
a source of ambiguities. [Actually, in their first paper, see the
Appendix A in Ref. \cite{JS98}, these authors did performed their
basic computation using the inconsistent rule ${\rm
Pf}(F\delta_1)=(F)_1{\rm Pf}\delta_1$. Later in Ref. \cite{DJS00a}
(see the Appendix A there), they argued that their result was
``stable'' against a possible violation of the latter rule.]  By
contrast, the authors
\cite{BF00,BFreg,BFregM,BFeom} have accepted the special features of
the partie finie, such as its non-distributivity, and constructed
by its mean a mathematically consistent regularization, able to give a
precise sense to all computations at the 3PN order.

The Hadamard partie finie $(F)_1$ of a singular function involves a
spherical average that is defined within the spatial hypersurface $t=$
const of a global coordinate system like the harmonic
coordinates. Clearly, this definition is incompatible with the
framework of a relativistic field theory, and we expect at some level
a violation of the Lorentz invariance of the equations of motion due
to this regularization. Remarkably, such a violation occurs only at
the 3PN order; up to the 2.5PN order the equations of motion in
harmonic coordinates, as computed using the regularization $(F)_1$,
are Lorentz-invariant \cite{BFP98}. To overcome this problem at the
3PN order, it has been necessary to define a ``Lorentzian''
regularization
\cite{BFregM}, which consists merely of applying the Hadamard partie
finie within the spatial hypersurface orthogonal to the (Minkowskian)
four-velocity of a particle. It was shown in Ref. \cite{BFeom} that
the Lorentzian regularization adds some new terms to the 3PN equations
of motion [computed with the standard regularization $(F)_1$] which
are mandatory in order to maintain their Lorentz invariance (see for
instance Eq. (5.35) in \cite{BFeom}). The Lorentzian partie finie of a
singular function $F$, denoted $[F]_1$, enables one to define a
``Lorentzian'' partie finie delta-function ${\rm Pf}\Delta_1$, namely
a linear form whose action on any $F$ gives the real number
$[F]_1$. It also permits the precise definition, given by Eq. (5.11)
in \cite{BFregM}, of a model for the stress-energy tensor of
point-particles in (post-Newtonian expansions of) general relativity.

Besides the compact-support integrals computed before, the equations
of motion contain many ``non-compact'' integrals, whose support
extends up to infinity and which are divergent at the location of the
particles. To them we assign systematically the value given by the
Hadamard partie-finie of a divergent integral: ${\rm Pf}\int d^3{\bf
x}~\!F$, see Eq. (3.1) in \cite{BFreg}. Furthermore, to any $F$ in
this class, we associate the pseudo-function ${\rm Pf}F$ which by
definition is the linear form whose action on any other $G$ gives the
real number ${\rm Pf}\int d^3{\bf x}~\!FG$. Given then two pseudo
functions their product is chosen to be the ``ordinary'' one ${\rm
Pf}F.{\rm Pf}G={\rm Pf}(FG)$.

An important feature of the Hadamard partie-finie integral is that the
integral of a gradient is not zero in general, ${\rm Pf}\int d^3{\bf
x}~\!\partial_iF\not= 0$, since it is equal to the sums of the parties
finies of the surface integrals surrounding the singularities when the
surface areas tend to zero; see Eq. (3.4) in \cite{BFreg}. This means
that the ordinary derivative of singular functions shows a fundamental
difference with the case of regular sources, since in this case the
integral of a gradient is always zero (provided that the integrand
decreases sufficiently fast at infinity). One can check that some
non-vanishing integrals of a gradient start to appear precisely at the
3PN order. Confronted with this problem, the authors
\cite{JS98,JS99,DJS00a,DJS00b,DJS00c} have considered that this signals 
the presence of ambiguities at the 3PN order, notably because their
ADM-Hamiltonian density is defined only modulo a total divergence,
that one certainly does not want to contribute even in the case of
singular sources. On the other hand, the authors
\cite{BF00,BFreg,BFregM,BFeom} have accepted this feature and introduced
a new kind of (spatial or temporal) distributional derivative acting
on the pseudo-functions of the type ${\rm Pf}F$ (for instance
$\partial_i{\rm Pf}F$) in order to ensure that the integral of a 
gradient is always zero. It was found
\cite{BFreg} that it is impossible to define a derivative which
satisfies the Leibniz rule for the derivation of a product,
i.e. $\partial_i({\rm Pf}FG)\not= F\partial_i{\rm Pf}G+G\partial_i{\rm
Pf}F$ in general, but that when one replaces the Leibniz rule by the
weaker rule of ``integration by parts'', an interesting mathematical
structure exists. By rule of integration by parts, we refer to the
relation $\int d^3{\bf x}~[F\partial_i{\rm Pf}G+G\partial_i{\rm
Pf}F]=0$, for $F$ and $G$ arbitrary functions (see Eq. (7.2) in
\cite{BFreg} where we use a more appropriate bracket notation for the
spatial integral). While the rule of integration by parts is nothing
but an integrated version of the ``pointwise'' Leibniz rule, the
Leibniz rule itself is a stronger requirement, which is not satisfied
in general as there are triplets of singular functions $F$, $G$, $H$
for which $\int d^3{\bf x}~H[F\partial_i{\rm Pf}G+G\partial_i{\rm
Pf}F]\not= \int d^3{\bf x}~H\partial_i({\rm Pf}FG)$. The motivation
for requiring the rule of integration by parts is that it is clearly
valid in the case of regular fluid systems. Notably it implies that
the integral of a gradient of any singular function of type $F$ is
zero. However, because it violates the Leibniz rule, the
distributional derivative cannot be completely satisfying on the
physical point of view.

Actually two different distributional derivatives, and therefore two
different regularizations, were introduced in Ref. \cite{BFreg}. A
``particular'' derivative, defined by Eq. (7.7) in \cite{BFreg}, was
first chosen for its simplicity. The two main properties of this
derivative are: that (i) it reduces to the ordinary derivative,
i.e. $\partial_i{\rm Pf}F={\rm Pf}(\partial_iF)$, whenever $F$ is
bounded near the singularities (in addition of being smooth everywhere
else), (ii) it obeys the rule of integration by parts. Though the
particular derivative is especially convenient to use in practical
computations, it does not follow from some ``unicity'' theorem. A more
interesting derivative, on the mathematical point of view, is the
so-called ``correct'' derivative (we follow the terminology of
Ref. \cite{BFeom}) which {\it does} satisfy a unicity theorem. Namely,
this derivative is obtained in Theorem 4 of \cite{BFreg} as the {\it
unique} derivative satisfying the properties (i) and (ii) above, and,
in addition, (iii) the rule of commutation of successive derivatives
(Schwarz lemma). As it turned out, the ``correct'' derivative, given
by Eq. (8.12) of \cite{BFreg}, depends on one arbitrary numerical
constant $K$. (Note that both the particular and correct derivatives
reduce to the derivative of the standard distribution theory
\cite{Schwartz} when applied on smooth test functions with compact
support.)

Summarizing, it is possible to construct a consistent regularization
based on the Hadamard partie finie, thus one can give a precise
meaning to any integral encountered in the computation, but there are
several possible prescriptions associated with different
distributional derivatives (and the Leibniz rule is not
satisfied). Our strategy has been to perform two computations of the
equations of motion, associated respectively with the ``particular''
and ``correct'' derivatives. Then the following was shown
\cite{BFeom}.

\medskip\noindent
(I) The 3PN equations of motion, when computed by means of the
Lorentzian regularization and the particular derivative, are in
agreement with the known equations of motion up to the 2.5PN order,
have the correct test-mass limit and most importantly are Lorentz
invariant (in a perturbative post-Newtonian sense).

\medskip\noindent
(II) Looking for the most general solution, allowed by the
regularization, for the 3PN equations of motion to admit a conserved
energy and a Lagrangian description, we find that they depend on two
unphysical gauge-constants $r'_1$ and $r'_2$ (associated with the
appearance of logarithms), and on one and only one physical constant
$\lambda$ which cannot be determined within the method. The equations
of motion possess all the physical properties that we expect, but the
presence of the unknown constant $\lambda$ is somewhat baffling, as it
probably reflects a physical incompleteness of the regularization.

\medskip\noindent
(III) When the correct distributional derivative is used instead of
the particular one, the equations of motion depend on $K$ in addition
to $r'_1$, $r'_2$ and $\lambda$. In this case we find that they are no
longer Lorentz invariant in general, but that there is a unique value
of $K$ for which the Lorentz invariance is recovered:
$K=\frac{41}{160}$. For this value the equations of motion have also
all the physical properties we expect.

\medskip\noindent
(IV) The different equations of motion as obtained by means of the
``particular'' and ``correct'' prescriptions (with $K=\frac{41}{160}$
in the second case) are {\it physically equivalent} in the sense that
they differ from each other by an infinitesimal change of
coordinates. This satisfying result indicates that the distributional
derivatives introduced in Ref. \cite{BFreg} constitute merely some
technical tools devoid of physical meaning.

\medskip
In the scenario (III) one may wonder why after having used the
Lorentzian regularization defined in Ref. \cite{BFregM} one still has
to adjust the constant $K$ to a certain value in order to get finally
the Lorentz invariance. The likely reason is that the distributional
derivatives we use (the particular and correct ones) have not been
defined in a Lorentz-invariant way, as their distributional terms are
made of the delta-pseudo-function ${\rm Pf}\delta_1$ instead of the
``Lorentzian'' delta-pseudo-function ${\rm Pf}\Delta_1$ (see
Eq. (3.36) in \cite{BFregM}). As a result, we find in the scenario
(III) that although most of the terms satisfy the requirement of
Lorentz invariance, notably the terms proportional to the combination
of masses $m_1^2m_2$ in the acceleration of particle one (these terms
are shown to behave correctly thanks to the Lorentzian
regularization), there still exists a limited class of terms,
proportional to $m_2^3$, that do not obey the Lorentz invariance
unless $K$ is adjusted to the value $\frac{41}{160}$. [In the scenario
(I) where there is no constant to adjust the latter terms behave
correctly.]

The problem of the Lorentz invariance of the equations of motion was
solved in a quite different way by the other group
\cite{JS98,JS99,DJS00a,DJS00b,DJS00c}. 
We recall that the harmonic-coordinate equations of motion are
manifestly Lorentz-invariant because the harmonic gauge condition
preserves the Poincar\'e symmetry. By contrast, the coordinate
conditions associated with the ADM Hamiltonian formalism do not
respect the Poincar\'e group, and therefore the authors
\cite{JS98,JS99,DJS00a,DJS00b,DJS00c} had to prove that their
Hamiltonian is compatible with the existence of generators in
phase-space such that the usual Poincar\'e algebra is satisfied.  More
precisely, they constructed a generic ``ambiguous'' dynamics at the
3PN order, parametrized by some unknown ambiguity parameters
associated notably with the non-distributivity of the Hadamard partie
finie and to the fact that the integral of a gradient, in an ordinary
sense, is not zero. They showed that there were only two ambiguity
parameters they denoted $\omega_{\rm kinetic}$ and $\omega_{\rm
static}$. (Actually, in the first paper \cite{JS98} they considered
only the ambiguity constant $\omega_{\rm kinetic}$ and obtained the
value $\omega_{\rm static}=\frac{1}{8}$. The static ambiguity was
introduced in the second paper \cite{JS99}.)  By imposing in an
{\it ad hoc} manner the existence of the Poincar\'e generators for
their ambiguous Hamiltonian, they showed \cite{DJS00b} that the
parameter $\omega_{\rm kinetic}$ is fixed uniquely to the value
$\frac{41}{24}$. This result was in fact obtained earlier \cite{BF00}
by comparing their expression of the energy of circular orbits
\cite{DJS00a} to the expression we got by means of the explicitly
Lorentz-invariant formalism described in the scenario (I)
above. Finally, having fixed $\omega_{\rm kinetic}$, there still remained
in the ADM-Hamiltonian formalism one and only one undetermined
constant $\omega_{\rm static}$, that we shall find to be equivalent,
in the sense of Eq. (\ref{1}), to the constant $\lambda$ appearing in
harmonic coordinates.  [Note that, despite the resemblance between the
value $K=\frac{41}{160}$ in the scenario (III) and the result
$\omega_{\rm kinetic}=\frac{41}{24}$, the constant $K$ can be fixed to
this unique value only if the sophisticated Lorentzian regularization
is used before. Without such a regularization, several other terms not
parametrized by $K$ would not behave correctly under Lorentz
transformations, and therefore no value of $K$ could be chosen in
order to restore the Lorentz invariance. In this sense the constant
$K$ is more ``specialized'' than the constant $\omega_{\rm kinetic}$.]
 
Finally, choosing one or the other of the two approaches advocated in
Refs. \cite{JS98,JS99,DJS00a,DJS00b,DJS00c} and
\cite{BF00,BFreg,BFregM,BFeom} for the 
regularization is a matter of taste. In view of the equivalence of the
final results, it is a good state of affairs that the two approaches
are different conceptually and technically.

\section{Theory}

\subsection{Noetherian conserved quantities for a generalized Lagrangian}

At the 1PN order, the equations of motion of two compact objects in
General Relativity, as derived in Refs. \cite{LD17,EIH}, can be
deduced from an ordinary Lagrangian, depending on the positions and
velocities of the bodies, which was obtained by Fichtenholz
\cite{F50}. At the next 2PN order, the equations of motion in harmonic
coordinates, as obtained in
\cite{DD81a,D82a,D82b}, can only be deduced from a ``generalized''
Lagrangian, depending not only on the positions and velocities but
also on the accelerations of the particles \cite{DD81a}. In
particular, this confirmed a result of Martin and Sanz \cite{MS79}
that $N$-body systems cannot admit an ordinary Lagrangian description
beyond the 1PN order, provided that the gauge conditions preserve the
Lorentz invariance (as it is the case for the harmonic
gauge). However, it has been shown by Damour and Sch\"afer \cite{DS85}
that there exists a special class of coordinates, which includes the
ones associated with the ADM formalism, such that the Lagrangian at
the 2PN order expressed by means of such coordinates becomes ordinary,
i.e. does not depend on accelerations anymore. This means that we can
eliminate the accelerations in the harmonic-coordinate Lagrangian at
the 2PN order by going to the ADM coordinates
\cite{DS85}. In this paper, we shall find that the 3PN terms in the
Lagrangian in harmonic coordinates depend also on accelerations, and
that, like at the 2PN order, these accelerations can be eliminated by
a suitable coordinate transformation to some ``pseudo-ADM''
coordinates, following the general method of redefinition of position
variables \cite{DS85,S84,BOC84,DS91}.

Strictly speaking, the dynamics of two compact bodies does not derive
from a Lagrangian at the 3PN approximation because of the radiation
reaction damping effect at the previous 2.5PN order. When speaking of
a 3PN Lagrangian or Hamiltonian, we always refer to the conservative
part of the dynamics, which corresponds to the ``even'' post-Newtonian
orders 1PN, 2PN and 3PN. As we shall see, the radiation reaction
effect manifests itself in the non-conservation at the 2.5PN
approximation of the conserved quantities associated with the
conservative 3PN dynamics [see Eqs. (\ref{29})].

Let us consider a harmonic-coordinate generalized
3PN Lagrangian

\begin{equation}\label{2}
L^{\rm harmonic}\equiv L[{\bf y}_A(t),{\bf v}_A(t),{\bf a}_A(t)] \;,
\end{equation}
depending on the instantaneous positions $y_A^i(t)\equiv {\bf y}_A(t)$
(with $A=1,2$ and $i=1,2,3$), coordinate velocities $v_A^i(t)\equiv
{\bf v}_A(t)=d{\bf y}_A/dt$, as well as coordinate accelerations
$a_A^i(t)\equiv {\bf a}_A(t)=d{\bf v}_A/dt$. Our harmonic-coordinate
3PN Lagrangian is given by (\ref{24}) below, but we do not need to be
so specific in the present Section, where most of the results hold in
fact for $N$-body systems ($A=1,\cdots ,N$). We assume that the
dependence of the Lagrangian (\ref{2}) upon the accelerations is
linear. As a matter of fact, it is always possible to eliminate from a
generalized post-Newtonian Lagrangian a contribution quadratic in the
accelerations by re-writing it in the form of a so-called
``double-zero'' term, which does not contribute to the equations of
motion, plus a term linear in the acceleration
\cite{DS85} (this argument can be extended to any term polynomial
in the accelerations).

The equations of motion of the $A$th body are deduced from the
Lagrangian by taking the functional derivative defined as

\begin{equation}\label{3}
\frac{\delta L}{\delta y_A^i} \equiv \frac{\partial L}{\partial y_A^i} 
- \frac{d}{dt} \bigg( \frac{\partial L}{\partial v_A^i} \bigg)+
\frac{d^2}{dt^2} \bigg(\frac{\partial L}{\partial a_A^i} \bigg)= 0\;. 
\end{equation}
We consider first, very generally, an infinitesimal transformation of
the path of the particle $A$ at some instant $t$, i.e. $\delta{\bf
y}_A(t)={\bf y}'_A(t)-{\bf y}_A(t)$. The corresponding variations of
its velocity and acceleration are $\delta{\bf v}_A(t)=d\delta{\bf
y}_A/dt$ and $\delta{\bf a}_A(t)=d\delta{\bf v}_A/dt$. Such a
transformation of the motion induces a variation of the Lagrangian,
namely $\delta L= L[{\bf y}'_A,{\bf v}'_A,{\bf a}'_A]-L[{\bf y}_A,{\bf
v}_A,{\bf a}_A]$ which is readily found to be expressible, at the
linearized order in $\delta{\bf y}_A$, in the form

\begin{equation} \label{4}
\delta L = \frac{d Q}{dt} +\sum_A \frac{\delta L}{\delta y_A^i} \delta y_A^i
+{\cal O}\left(\delta{\bf y}_A^2\right)\;,
\end{equation}
where the functional derivative ${\delta L}/{\delta y_A^i}$ is given
by (\ref{3}) [it is zero ``on shell'', i.e. when the equations of
motion are satisfied], and where we have introduced the total
time-derivative of a function $Q\equiv Q[\delta {\bf y}_A,\delta {\bf
v}_A]$ defined by

\begin{equation}\label{5}
Q = \sum_A \left(p_A^i \delta y_A^i+q_A^i \delta v_A^i\right)\;.
\end{equation}
Here, $p_A^i$ and $q_A^i$ denote the momenta that are conjugate to
the positions $y_A^i$ and velocities $v_A^i$ of the
particle $A$ respectively, that is

\begin{mathletters}\label{6}\begin{eqnarray}
p_A^i &=& \frac{\delta L}{\delta v_A^i}\equiv \frac{\partial
L}{\partial v_A^i}-\frac{d}{dt} \bigg( \frac{\partial L}{\partial
a_A^i} \bigg)\;, \label{6a}\\ q_A^i &=& \frac{\delta L}{\delta
a_A^i}\equiv \frac{\partial L}{\partial a_A^i}\;.\label{6b}
\end{eqnarray}\end{mathletters}$\!\!$

We now discuss the Noetherian conservation laws for generalized
Lagrangians following Refs. \cite{DD81b,D82b}. We know from
Ref. \cite{BFeom} that the 3PN equations of motion in harmonic coordinates
are manifestly invariant (in a perturbative post-Newtonian sense)
under the Lorentz and more generally the Poincar\'e group. Thus the
dynamics associated with our 3PN generalized Lagrangian (\ref{24})
should stay the same after an infinitesimal Poincar\'e transformation
of the dynamical variables $y_A^\mu=(ct,{\bf y}_A)$. In particular,
this means that $\delta L=0$ in the case of arbitrary infinitesimal
constant spatial translations and rotations, $\delta y_A^i=\epsilon^i$
and $\delta y_A^i=\omega^i_{~j}y_A^j$ with
$\omega_{ij}=-\omega_{ji}$. In this case Eq. (\ref{4}) implies the
conservation on-shell (all the $\delta L/\delta y_A^i$'s are zero) of
the Noetherian linear and angular momenta given by
 
\begin{mathletters}\label{7}\begin{eqnarray}
P^i&=&\sum_A p_A^i\;,\label{7a}\\ J^i&=&\varepsilon_{ijk}\sum_A
\left(y_A^jp_A^k+v_A^jq_A^k\right)\;.\label{7b}
\end{eqnarray}\end{mathletters}$\!\!$
Thus, $dP^i/dt=0$ and $dJ^i/dt=0$ on shell.  On the other hand, we
have $\delta L=\tau dL/dt$ in the case of an infinitesimal constant
time translation $\delta t=\tau$, hence the conservation on-shell of
the Noetherian energy from Eq. (\ref{4}),

\begin{equation}\label{8}
E=\sum_A\left(v_A^ip_A^i+a_A^iq_A^i\right)-L\;.
\end{equation}
Thus, $dE/dt=0$. We shall give the explicit expressions of these
Noetherian energy and momenta at the 3PN order in harmonic coordinates
in the next Section which is devoted to the results [see
Eqs. (\ref{25})-(\ref{27})].

Finally, let us consider the symmetry of the Lagrangian that is
associated with the invariance under Lorentz special transformations
or boosts. Clearly, since the dynamics must stay the same after an
infinitesimal constant Lorentz boost, the corresponding variation of
the Lagrangian has to take essentially the form of a total time
derivative.  At the linearized order in the boost velocity $W^i$, the
transformation of the particle trajectories is given by $\delta
y_A^i=-W^i t +\frac{1}{c^2}W^jy_A^j v_A^i+{\cal O}(W^iW^i)$. There
should exist a certain functional $Z^i$ of the positions, velocities
and accelerations such that the 3PN Lagrangian variation reads $\delta
L=W^i dZ^i/dt+{\cal O}(W^iW^i)$, plus some ``double-zero'' terms at
the 3PN order (which are zero on-shell when applying the Noether
theorem). By applying Eq. (\ref{4}), we readily find the conservation
on-shell of the Noetherian integral $K^i=G^i-P^i t$, where $P^i$ is
the linear momentum (\ref{7a}), and where $G^i$ represents the
center-of-mass position:

\begin{equation}\label{9}
G^i=-Z^i+\sum_A\left(-q_A^i+\frac{1}{c^2}\left[y_A^ip_A^jv_A^j
+y_A^iq_A^ja_A^j+v_A^iq_A^jv_A^j\right]\right)\;.
\end{equation}
Thus, $dK^i/dt=0$, or equivalently $d^2G^i/dt^2=0$ (the center-of-mass
vector $G^i$ is conserved in a frame where $P^i=0$). The existence of
the latter boost-symmetry of the Lagrangian is a confirmation of the
Lorentz invariance of the 3PN equations of motion obtained in
Ref. \cite{BFeom}.  The Noetherian center-of-mass $G^i$ in harmonic
coordinates at the 3PN order is given explicitly by Eq. (\ref{28})
below.
 
The ten Noetherian quantities (\ref{7})-(\ref{9}) have been found from
our generalized Lagrangian as some functionals of the positions,
velocities and accelerations of the particles. However, once they have
been constructed, all the accelerations they involve can be
order-reduced using the fact that they take on-shell some definite
expressions depending on the positions and velocities as given by the
equations of motion. Our final results presented in Section IV.A have
all been order-reduced consistently with the 3PN approximation.

\subsection{Elimination of acceleration-dependent terms in a Lagrangian}

We start from the harmonic coordinate system $x^\mu=(ct,{\bf x})$ and
perform an infinitesimal coordinate transformation to a new coordinate
system ${x'}^\mu$, generally not obeying the harmonic gauge condition,
of the type

\begin{equation}\label{10}
{x'}^\mu=x^\mu+\varepsilon^\mu (x)\;,
\end{equation}
where $\varepsilon^\mu (x)$ is a function of the spatial coordinates
${\bf x}$ as well as a (local-in-time) functional of the trajectories
${\bf y}_A(t)$ and velocities ${\bf v}_A(t)$ parametrized by the
coordinate time $t=x^0/c$. Namely,

\begin{equation}\label{11}
\varepsilon^\mu ({\bf x},t)=\varepsilon^\mu [{\bf x};{\bf y}_A(t),
{\bf v}_A(t)]\;.
\end{equation}
Since the accelerations in the harmonic-coordinate Lagrangian appear
only at the 2PN order, we suppose that the coordinate transformation
starts at the same level. This means that $\varepsilon^i={\cal
O}\left(\frac{1}{c^4}\right)$ and $\varepsilon^0={\cal
O}\left(\frac{1}{c^3}\right)$. In particular we can check that any
term in the following which is at least quadratic in $\varepsilon^\mu$
is in fact of order ${\cal O}\left(\frac{1}{c^8}\right)$ and thus can
be neglected in our study limited to the 3PN approximation.  The
trajectories and velocities in the new coordinates ${x'}^\mu=(ct',{\bf
x}')$ are some functions ${\bf y}'_A(t')$ and ${\bf v}'_A(t')$ of the
new coordinate time $t'={x'}^0/c$.  The ``contact'' transformation of
the particle variables induced by the coordinate transformation
(\ref{10})-(\ref{11}) is defined by $\delta
y_A^i(t)={y'}_A^i(t)-y_A^i(t)$ (we use the same terminology as in
Ref. \cite{DS85}). Neglecting all the terms of the order of the square
of $\varepsilon^\mu$ we obtain

\begin{equation}\label{12}
\delta y_A^i(t)=\varepsilon^i({\bf y}_A,t)-\frac{v_A^i}{c}
\varepsilon^0({\bf y}_A,t)+{\cal O}
\left(\frac{1}{c^8}\right)\;.
\end{equation}
In this paper we shall construct a contact transformation $\delta
y_A^i$, composed of 2PN and 3PN terms and neglecting ${\cal
O}\left(\frac{1}{c^8}\right)$, which is issued from some infinitesimal
coordinate transformation (\ref{10})-(\ref{11}); however we shall not
be so much interested in the coordinate transformation itself, in
particular this means that we shall not investigate to which
coordinate conditions it corresponds to (non-harmonic and/or
ADM-type).

If the equations satisfied by the world-lines ${\bf y}_A(t)$ in some
initial coordinate system derive from the Lagrangian $L$, then the
equations satisfied by the new world-lines ${\bf y}'_A(t')$ in a new
coordinate system will derive from the new Lagrangian $L'$ that is
such that

\begin{equation}\label{13}
L'[{\bf y}'_A(t),{\bf v}'_A(t),{\bf a}'_A(t),{\bf b}'_A(t)]=L[{\bf
y}_A(t),{\bf v}_A(t),{\bf a}_A(t)]
\end{equation}
(see e.g. Eq. (5) of Damour and Sch\"afer \cite{DS85}). Since we
assumed that the contact transformation $\delta{\bf y}_A$ depends on
the velocities, the new Lagrangian necessarily depends on positions,
velocities, accelerations and also derivatives of accelerations: ${\bf
b}_A(t)=d{\bf a}_A/dt$. Now the same computation as the one leading to
Eq. (\ref{4}) shows that, at the linearized order in $\delta{\bf
y}_A$,

\begin{equation}\label{14}
L'[{\bf y}_A,{\bf v}_A,{\bf a}_A,{\bf b}_A]=L[{\bf y}_A,{\bf v}_A,{\bf
a}_A]+\frac{d Q}{dt}+\sum_A\frac{\delta L}{\delta y_A^i}\delta
y_A^i+{\cal O}
\left(\frac{1}{c^8}\right)\;.
\end{equation}
Notice that both sides of this relation are expressed in terms of the
same ``dummy'' variables, chosen to be the harmonic-coordinate ones,
e.g. ${\bf y}_A$. At the end, when we obtain the new Lagrangian,
we shall have to replace this dummy variable by the one corresponding
to the new coordinate system, ${\bf y}'_A={\bf y}_A+\delta{\bf
y}_A$. The term with a total time-derivative is the same as the one
found in Eq. (\ref{4}), with $Q$ given by (\ref{5}). As one can see,
the dependence of the Lagrangian $L'$ upon derivatives of
accelerations ${\bf b}_A$ comes only from this total time
derivative. Therefore, by posing $L''=L'-\frac{d Q}{dt}$ we get a
Lagrangian which is dynamically equivalent to the Lagrangian $L'$ and
depends like $L$ on positions, velocities and accelerations only,

\begin{equation}\label{15}
L''[{\bf y}_A,{\bf v}_A,{\bf a}_A]=L[{\bf y}_A,{\bf v}_A,{\bf
a}_A]+\sum_A\frac{\delta L}{\delta y_A^i}\delta y_A^i+{\cal O}
\left(\frac{1}{c^8}\right)\;.
\end{equation}

We now show that there exists a contact transformation $\delta y_A^i$
(actually, there exist infinitely many of them), together with a
redefinition of the Lagrangian by the addition of a total time
derivative, which eliminates all the accelerations in the Lagrangian
up to the 3PN order. In other words, the 3PN Lagrangian that will
follow is ordinary, i.e. depends on positions and velocities
only. Damour and Sch\"afer
\cite{DS85} have already shown how to eliminate the accelerations at
the 2PN level. We shall see how to do this at the next 3PN order, but
in fact the method is a particular application of a general algorithm
to eliminate higher-derivative terms in a Lagrangian
\cite{DS91}. Since the contact transformation (\ref{12}) is assumed to
start at the 2PN order, i.e. $\delta y_A^i={\cal
O}\left(\frac{1}{c^4}\right)$, we must control the functional
derivative $\frac{\delta L}{\delta y_A^i}$ appearing in the right side
of Eq. (\ref{15}) at the relative 1PN order. The standard Newtonian
contribution is then followed by a certain 1PN correction, denoted
$m_AC_A^i$, hence

\begin{equation}\label{16}
\frac{\delta L}{\delta y_A^i}=m_A\Bigg[- a_A^i-\sum_{B\not= A}
\frac{G m_B}{r_{AB}^2}n_{AB}^i+\frac{1}{c^2}C_A^i\Bigg]+{\cal O}
\left(\frac{1}{c^4}\right)\;.
\end{equation}
The 1PN term $C_A^i$ can be straightforwardly computed from the
Lagrangian (\ref{24}). The point is that it does depend on
accelerations, $C_A^i\equiv C_A^i[{\bf y}_B,{\bf v}_B,{\bf a}_B]$,
with this dependence being {\it linear}. The presence of accelerations
in $C_A^i$ is the reason why the method used in Ref. \cite{DS85} to
deal with the problem at the 2PN order cannot be extended immediately
at the 3PN approximation. We shall see that the method necessitates
the introduction in the contact transformation at the 3PN order of
some ``counter-term'' $X_A^i$ described below. Now, in view of the
term $-m_Aa_A^i$ present in Eq. (\ref{16}), it is clear that we will
be able to remove all the accelerations at the 2PN order if we choose
for the contact transformation the term $\frac{1}{m_A}q_A^i$ (we
recall that $q_A^i$ is the conjugate momentum of the acceleration,
$q_A^i=\frac{\partial L}{\partial a_A^i}$). Indeed the only possible
accelerations at the 2PN order in the Lagrangian $L''$ would be
contained in the combination $L-\sum_A a_A^iq_A^i$, which clearly does
{\it not} depend on accelerations because of the linearity of the
original Lagrangian $L$ upon $a_A^i$. Furthermore, as discussed in
\cite{DS85}, once we have eliminated the 
accelerations at the 2PN order, we are free to add to the contact
transformation any term of the type $\frac{1}{m_A}\frac{\partial
F}{\partial v_A^i}$, where $F$ is an arbitrary functional of the
positions and velocities only, starting at the 2PN order. This follows
immediately from the identity $\frac{dF}{dt}=\sum_A
\big(v_A^i\frac{\partial F}{\partial y_A^i}+ a_A^i\frac{\partial
F}{\partial v_A^i}\big)$, which shows that the further accelerations
produced by this term are contained into the total time-derivative of
$F$, and so can be removed from the original Lagrangian without
changing the dynamics. However, these procedures are no longer valid
at the 3PN order because of the accelerations in the 1PN term $C_A^i$
of (\ref{16}), which will couple to the terms
$\frac{1}{m_A}\left[q_A^i+\frac{\partial F}{\partial v_A^i}\right]$ as
suggested before and produce some new accelerations. The solution of
the problem is to add to the contact transformation some correction
term that we shall find to be adjustable in a unique way so that it
works.

As a result, we look for a contact transformation of the type

\begin{equation}\label{17}
\delta y_A^i=\frac{1}{m_A}\left[q_A^i+\frac{\partial F}{\partial v_A^i}
+\frac{1}{c^6}X_A^i\right]+{\cal O}\left(\frac{1}{c^8}\right)\;,
\end{equation}
where $q_A^i$ is defined by Eq. (\ref{6b}); $F$ is a general
functional of the positions and velocities, $F\equiv F[{\bf y}_A,{\bf
v}_A]$, and $X_A^i$ denotes some ``counter'' term depending on
positions and velocities only, $X_A^i\equiv X_A^i[{\bf y}_B,{\bf
v}_B]$. We recall that $q_A^i$ is composed of 2PN and 3PN terms, which
are easily computed from the Lagrangian (\ref{24}). The function $F$
must start at the 2PN order; in addition we assume that it contains
all possible generic terms at 3PN. Finally as explained above the
counter term $X_A^i$ is purely of order 3PN. We now replace both
Eqs. (\ref{16}) and (\ref{17}) into $L''$ given by (\ref{15}) and
investigate the occurence of accelerations. Among the terms we
recognize the combination $L-\sum_A a_A^iq_A^i$ which is free of any
accelerations at the 3PN order. We also transfer several acceleration
terms into the total time-derivative of $F$ as before. At last we find
that the only remaining accelerations in $L''$ are contained into the
particular combination of terms:

$$\sum_A\left(\frac{1}{c^2}\left[q_A^i+\frac{\partial F}{\partial
v_A^i}\right]C_A^i -\frac{1}{c^6}a_A^iX_A^i\right)+{\cal O}
\left(\frac{1}{c^8}\right)\;.$$ 
As all the terms in that combination are linear in the accelerations,
we see that for {\it any} given function $F$ there is a {\it unique}
choice of the term $X_A^i$ (for each particle) such that all the
remaining accelerations are cancelled out, namely

\begin{equation}\label{19}
\frac{1}{c^6}X_A^i=\sum_B\frac{1}{c^2}\left[q_B^j+\frac{\partial F}
{\partial v_B^j}\right]\frac{\partial C_B^j}{\partial a_A^i}+{\cal O}
\left(\frac{1}{c^8}\right)\;.
\end{equation}
With the latter choice, the contact transformation (\ref{17}), defined
for any $F$, yields a Lagrangian $L''$ whose only accelerations come
from (minus) the total time-derivative of $F$. Therefore, the 3PN
Lagrangian $L'''=L''+\frac{dF}{dt}$ is at once physically equivalent
to $L''$, $L'$ and $L$, and free of accelerations. Our result reads
then

\begin{equation}\label{20}
L'''[{\bf y}_A,{\bf v}_A]=L+\sum_A\frac{\delta L}{\delta y_A^i}\delta
y_A^i+\frac{dF}{dt}+{\cal O}\left(\frac{1}{c^8}\right)\;.
\end{equation}
Remind the large freedom we still have on the definition of $L'''$,
since we constructed it for any functional $F$ of the positions and
velocities at the 2PN and 3PN orders.

In this paper we shall be able to determine uniquely the function $F$
by the requirement that the Lagrangian $L'''$ be exactly the ADM
Lagrangian associated with the ADM (or ADM-type) Hamiltonian
published by Damour, Jaranowski and Sch\"afer
\cite{DJS00b}. We shall not give the details of the computation since
it consists merely of parametrizing the most general function $F$,
constructed with the dynamical variables of the problem and having a
compatible dimension, by means of some arbitrary constant parameters,
and to show that all these constants are uniquely fixed by the
condition of matching to the ADM Hamiltonian. We find indeed, in
complete agreement with Ref. \cite{DJS00c}, that there is a unique set
of constants for which this works. In particular the equivalence is
possible if and only if the undetermined constant $\lambda$ appearing
in the harmonic-coordinate formalism
\cite{BFeom} is related to the constant $\omega_{\rm static}$ of
Jaranowski and Sch\"afer \cite{JS99} by Eq. (\ref{1}). Note that the
latter matching shows also that the logarithms $\ln
\big(\frac{r_{12}}{r'_1}\big)$ and $\ln \big(\frac{r_{12}}{r'_2}\big)$
present in the harmonic-coordinate Lagrangian (\ref{24}), where $r'_1$
and $r'_2$ denote some regularization constants, are eliminated by
this contact transformation, in agreement with the fact proved in
Ref. \cite{BFeom} that the logarithms, and the constants $r'_1$ and
$r'_2$ therein, can be gauged away. See Eq. (\ref{31}) below for the
complete expression of the function $F$.

At last, with $F$ now fully specified by the equivalence with
\cite{DJS00b}, we obtain the ordinary ADM-type Lagrangian

\begin{equation}\label{21}
L^{\rm ADM}=L+\sum_A\frac{\delta L}{\delta y_A^i}\delta
y_A^i+\frac{dF}{dt}\;,
\end{equation}
given explicitly at the 3PN order by Eq. (\ref{33}) below, in which,
as mentionned above, we shall replace the ``dummy'' variables used in
the computation, $y_A^i$ and $v_A^i$, by the real dynamical variables
in pseudo-ADM coordinates, $Y_A^i$ and $V_A^i$. The ADM momentum
conjugate to the velocity is

\begin{equation}\label{22}
P_A^i=\frac{\partial L^{\rm ADM}}{\partial v_A^i}=p_A^i+
\frac{\delta}{\delta v_A^i}\left(\sum_B\delta
y_B^j\frac{\delta L}{\delta
y_B^j}\right)+\frac{\partial F}{\partial y_A^i}\;,
\end{equation}
and the corresponding Hamiltonian follows from the ordinary Legendre
transformation

\begin{equation}\label{23}
H^{\rm ADM}=\sum_A P_A^i v_A^i-L^{\rm ADM}\;.
\end{equation}
See Eq. (\ref{34}) for the complete 3PN expression of this Hamiltonian
(as a function of $Y_A^i$ and $P_A^i$). [We have checked that the
second equality in (\ref{22}) is true at 3PN order.] Notice that,
strictly speaking, $H^{\rm ADM}$ is not the ADM one, as it differs
from it by a shift in phase-space coordinates at the 3PN order which
is given in Ref.
\cite{DJS00b}. Indeed, the ADM Hamiltonian at the 3PN order is not
ordinary, as it depends on the positions and momenta as well as on
their derivatives \cite{JS98}. But this is not a concern for our
purpose, since we are interested in proving the equivalence between
our approach \cite{BF00,BFreg,BFregM,BFeom} and the one of
\cite{JS98,JS99,DJS00a,DJS00b,DJS00c}, that is in finding the existence of a
unique transformation connecting both works, in whatever coordinate
systems the two approaches found it convenient to be. We think that
the equivalence found in this paper and in Ref. \cite{DJS00c}
convincingly confirms the correctness of the result. This equivalence
is especially important in view of the different procedures adopted by
the two groups to treat the point-mass divergencies (see Section II
for a discussion).

\section{Results}

\subsection{Conserved quantities in harmonic coordinates at the 3PN order}

We first exhibit a generalized Lagrangian from which derive the 3PN
equations of motion of two compact objects as they were obtained in
harmonic coordinates; see Eqs. (7.16) in \cite{BFeom}. The Lagrangian
corresponds only to the conservative part of the equations, which
excludes the radiation reaction term at the 2.5PN order. To compute it
we proceed by guess-work, and find the occurence of terms depending on
accelerations at the 2PN and 3PN orders. The Lagrangian is chosen to
be linear in the accelerations, and to agree at the 2PN approximation
with the Lagrangian obtained in Ref. \cite{D82b}. The result is

\begin{align}\label{24}
 L & = 
\frac{G m_1 m_2}{2 r_{12}} + \frac{m_1 v_1^2}{2} \nonumber \\ & + 
\frac{1}{c^2} \Bigg\{-\frac{G^2 m_1^2 m_2}{2 r_{12}^2} + \frac{m_1 v_1^4}{8 }+
\frac{G m_1 m_2}{r_{12}} \bigg(-\frac{1}{4} (n_{12}v_1) (n_{12}v_2) +
\frac{3}{2} v_1^2 - \frac{7}{4} 
(v_1v_2)\bigg)\Bigg\} \nonumber \\ & + 
\frac{1}{c^4} \Bigg\{\frac{G^3 m_1^3 m_2}{2 r_{12}^3}+ \frac{19 G^3 m_1^2
m_2^2}{8 r_{12}^3} + \frac{G^2 m_1^2 m_2}{r_{12}^2} \bigg(\frac{7}{2}
(n_{12}v_1)^2 - \frac{7}{2} (n_{12}v_1) (n_{12}v_2) +
\frac{1}{2}(n_{12}v_2)^2
\nonumber \\ & \qquad  + \frac{1}{4} v_1^2 - \frac{7}{4} (v_1v_2) +
\frac{7}{4} v_2^2\bigg) + \frac{G m_1 m_2}{r_{12}} \bigg(\frac{3}{16}
(n_{12}v_1)^2 (n_{12}v_2)^2 - \frac{7}{8} (n_{12}v_2)^2 v_1^2 + \frac{7}{8}
v_1^4 \nonumber \\ & \qquad + \frac{3}{4} (n_{12}v_1) (n_{12}v_2) (v_1v_2) - 2
v_1^2 (v_1v_2) + \frac{1}{8} 
(v_1v_2)^2 + \frac{15}{16} v_1^2 v_2^2\bigg) + \frac{m_1 v_1^6}{16}
\nonumber \\ & \qquad +
G m_1 m_2 \bigg(-\frac{7}{4}
(a_1 v_2) (n_{12}v_2) - \frac{1}{8} (n_{12} a_1) (n_{12}v_2)^2 + \frac{7}{8}
(n_{12} a_1) v_2^2\bigg)\Bigg\} \nonumber \\ & + 
\frac{1}{c^6} \Bigg\{\frac{G^2 m_1^2 m_2}{r_{12}^2} \bigg(\frac{13}{18}
(n_{12}v_1)^4 + \frac{83}{18} (n_{12}v_1)^3 
(n_{12}v_2) - \frac{35}{6} (n_{12}v_1)^2 (n_{12}v_2)^2 - \frac{245}{24}
(n_{12}v_1)^2 v_1^2 \nonumber \\ & \qquad + \frac{179}{12} (n_{12}v_1)
(n_{12}v_2) v_1^2 
- \frac{235}{24} (n_{12}v_2)^2 v_1^2 + \frac{373}{48} v_1^4 + \frac{529}{24}
(n_{12}v_1)^2 (v_1v_2) \nonumber \\ & \qquad - \frac{97}{6} (n_{12}v_1)
(n_{12}v_2) 
(v_1v_2) - \frac{719}{24} v_1^2 (v_1v_2) + \frac{463}{24} (v_1v_2)^2 -
\frac{7}{24} (n_{12}v_1)^2 v_2^2 \nonumber \\ & \qquad - \frac{1}{2}
(n_{12}v_1) 
(n_{12}v_2) v_2^2 + \frac{1}{4} 
(n_{12}v_2)^2 v_2^2 + \frac{463}{48} v_1^2 v_2^2 - \frac{19}{2} (v_1v_2) v_2^2
+ \frac{45}{16} v_2^4\bigg) +  \frac{5m_1 v_1^8}{128}  \nonumber \\ & \qquad
+ G m_1 
m_2 \bigg(\frac{3}{8} (a_1 v_2) (n_{12}v_1) (n_{12}v_2)^2 + \frac{5}{12} (a_1
v_2) (n_{12}v_2)^3 + \frac{1}{8} (n_{12} a_1) (n_{12}v_1) (n_{12}v_2)^3
\nonumber \\ &  
\qquad + \frac{1}{16} (n_{12} a_1) (n_{12}v_2)^4 +
\frac{11}{4} (a_1 v_1) (n_{12}v_2) v_1^2 - (a_1 v_2) (n_{12}v_2) v_1^2 - 2
(a_1 v_1) 
(n_{12}v_2) (v_1v_2) \nonumber \\ & \qquad + \frac{1}{4} (a_1 v_2) (n_{12}v_2)
(v_1v_2) + 
\frac{3}{8} (n_{12} a_1) (n_{12}v_2)^2 (v_1v_2) - \frac{5}{8} (n_{12} a_1)
(n_{12}v_1)^2 v_2^2 \nonumber \\ & \qquad + \frac{15}{8} (a_1 v_1) (n_{12}v_2)
v_2^2 - \frac{15}{8} (a_1 v_2) (n_{12}v_2) v_2^2 - 
\frac{1}{2} (n_{12} a_1) (n_{12}v_1) (n_{12}v_2) v_2^2 \nonumber \\ & \qquad -
\frac{5}{16} (n_{12} a_1) (n_{12}v_2)^2 v_2^2\bigg) +
\frac{G^2 m_1^2 m_2}{r_{12}} \bigg(-\frac{235}{24} (a_2 v_1) (n_{12}v_1) -
\frac{29}{24} (n_{12} 
a_2) (n_{12}v_1)^2 \nonumber \\ & \qquad - \frac{235}{24} (a_1 v_2)
(n_{12}v_2) - \frac{17}{6} 
(n_{12} a_1) (n_{12}v_2)^2 + \frac{185}{16} (n_{12} a_1) v_1^2 -
\frac{235}{48} (n_{12} a_2) v_1^2 \nonumber \\ & \qquad -
\frac{185}{8} (n_{12} a_1) (v_1v_2) + \frac{20}{3} (n_{12} a_1) v_2^2\bigg) +
\frac{G 
m_1 m_2}{r_{12}} \bigg(-\frac{5}{32} (n_{12}v_1)^3 (n_{12}v_2)^3 \nonumber \\
& \qquad + 
\frac{1}{8} (n_{12}v_1) (n_{12}v_2)^3 v_1^2 +
\frac{5}{8} (n_{12}v_2)^4 v_1^2 - \frac{11}{16} (n_{12}v_1) (n_{12}v_2) v_1^4
+ \frac{1}{4} (n_{12}v_2)^2 v_1^4 + \frac{11}{16} v_1^6 \nonumber \\ & \qquad 
- \frac{15}{32}
(n_{12}v_1)^2 (n_{12}v_2)^2 (v_1v_2) + (n_{12}v_1) (n_{12}v_2) v_1^2 (v_1v_2)
+ \frac{3}{8} (n_{12}v_2)^2 v_1^2 (v_1v_2) \nonumber \\ & \qquad -
\frac{13}{16} v_1^4 (v_1v_2) + 
\frac{5}{16} (n_{12}v_1) (n_{12}v_2) (v_1v_2)^2 + \frac{1}{16} (v_1v_2)^3 -
\frac{5}{8} (n_{12}v_1)^2 v_1^2 v_2^2 \nonumber \\ & \qquad - \frac{23}{32}
(n_{12}v_1) 
(n_{12}v_2) v_1^2 v_2^2 + \frac{1}{16} v_1^4
v_2^2 - \frac{1}{32} v_1^2 (v_1v_2) v_2^2\bigg) -\frac{3 G^4 m_1^4 m_2}{8
r_{12}^4 } \nonumber \\ & \qquad + \frac{G^4 m_1^3 m_2^2}{r_{12}^4}
\bigg(-\frac{5809}{280}  +
\frac{11}{3} \lambda + \frac{22}{3} \ln 
\left(\frac{r_{12}}{r'_1} \right)\bigg) + \frac{G^3 m_1^2 m_2^2}{r_{12}^3}
\bigg(\frac{383}{24} 
(n_{12}v_1)^2 \nonumber \\ & \qquad - \frac{889}{48} (n_{12}v_1) (n_{12}v_2) -
\frac{123}{64} 
(n_{12}v_1)^2 \pi^2 + \frac{123}{64} (n_{12}v_1) (n_{12}v_2) \pi^2 -
\frac{305}{72} v_1^2 + \frac{41}{64} \pi^2 v_1^2 \nonumber \\ & \qquad +
\frac{439}{144} 
(v_1v_2) - \frac{41}{64} \pi^2 (v_1v_2)\bigg) + \frac{G^3 m_1^3 m_2}{r_{12}^3}
\bigg(-\frac{8243}{210} (n_{12}v_1)^2  +
\frac{15541}{420} (n_{12}v_1) (n_{12}v_2) \nonumber \\ & \qquad + \frac{3}{2}
(n_{12}v_2)^2 +
\frac{15611}{1260} v_1^2 - \frac{17501}{1260} (v_1v_2) + \frac{5}{4} v_2^2 
\nonumber \\ & \qquad +
22 (n_{12}v_1)^2 \ln \left(\frac{r_{12}}{r'_1} \right) - 22
(n_{12}v_1) (n_{12}v_2) \ln \left(\frac{r_{12}}{r'_1} \right) -
\frac{22}{3} v_1^2 \ln \left(\frac{r_{12}}{r'_1} \right) \nonumber \\
& \qquad + \frac{22}{3} (v_1v_2) \ln
\left(\frac{r_{12}}{r'_1} \right)\bigg)\Bigg\}+1 \leftrightarrow 2 +{\cal
O}\left(\frac{1}{c^7}\right)\;.
\end{align}
In our notation, $r_{12}=|{\bf y}_1 - {\bf y}_2|$, ${\bf n}_{12}=({\bf
y}_1-{\bf y}_2)/r_{12}$, and the scalar products are written
e.g. $(n_{12}v_2)={\bf n}_{12}.{\bf v}_2$. To the terms given
explicitly above, we have to add the terms corresponding to the
relabeling $1 \leftrightarrow 2$, including those which are symmetric
under the label exchange. Notice the presence of the constant
$\lambda$ which is the only unknown physical parameter in this
Lagrangian, and of the two unknown gauge constants $r'_1$ and $r'_2$
(we follow exactly the notation of
\cite{BFeom}). The Lagrangian presented here
is not the only admissible one, as we can always add to it an
arbitrary total time derivative (double-zero terms would make the
Lagrangian non-linear in the accelerations). We have checked that our
Lagrangian (\ref{24}) differs indeed from the one given by Eqs. (5.4)-(5.10)
in Ref. \cite{DJS00c} by a total time derivative.

Next we present the expressions of the conserved integrals of the 3PN
harmonic-coordinate motion as constructed in Section III.A. These
expressions involve only the relativistic 1PN, 2PN and 3PN terms
corresponding to the conservative part of the dynamics at the 3PN
order. The radiation reaction damping effect is added afterwards. All
the quantities we present depend only on the positions and velocities,
because all accelerations therein have been systematically
order-reduced by means of the equations of motion. The energy $E$
reads

\begin{eqnarray}
E &=& \frac{ m_1 v_1^2}{2 } -\frac{G m_1 m_2}{2 r_{12}} \nonumber\\ &+&  
\frac{1}{c^2} \Bigg\{\frac{G^2 m_1^2 m_2}{2 r_{12}^2 } + \frac{3  
m_1 v_1^4}{8 }+ \frac{G m_1 m_2}{r_{12}}
\bigg(-\frac{1}{4}(n_{12}v_1) (n_{12}v_2) + \frac{3}{2} v_1^2  -
\frac{7}{4} (v_1v_2) \bigg)\Bigg\} \nonumber\\ &+&
\frac{1}{c^4}\Bigg\{-\frac{G^3 m_1^3 m_2}{2 r_{12}^3}- \frac{19 G^3
m_1^2 m_2^2}{8 r_{12}^3} + \frac{5 m_1 v_1^6}{16} +
\frac{G m_1 m_2}{r_{12}} \bigg(\frac{3}{8} (n_{12}v_1)^3 
(n_{12}v_2)\nonumber\\ & & \quad + \frac{3}{16} (n_{12}v_1)^2 (n_{12}v_2)^2
- \frac{9}{8} 
(n_{12}v_1) (n_{12}v_2) v_1^2 - 
 \frac{13}{8} (n_{12}v_2)^2 v_1^2 + \frac{21}{8} v_1^4  \nonumber\\ & & \quad
+ 
\frac{13}{8}   (n_{12}v_1)^2 (v_1v_2) + \frac{3}{4} (n_{12}v_1) (n_{12}v_2)
(v_1v_2) - \frac{55}{8} v_1^2 (v_1v_2) +
\frac{17}{8} (v_1v_2)^2  + \frac{31}{16} v_1^2 v_2^2 \bigg) \nonumber\\ & &
\quad + 
\frac{G^2 m_1^2 m_2}{r_{12}^2} 
\bigg(\frac{29}{4} (n_{12}v_1)^2 - \frac{13}{4} (n_{12}v_1) (n_{12}v_2) +
\frac{1}{2}(n_{12}v_2)^2 - \frac{3}{2} v_1^2 + \frac{7}{4} v_2^2\bigg)\Bigg\}
\nonumber\\ &+& \frac{1}{c^6} \Bigg\{\frac{35  m_1 v_1^8}{128 }+ \frac{G m_1
m_2}{ 
r_{12}}  \bigg(-\frac{5}{16} (n_{12}v_1)^5 (n_{12}v_2) 
- \frac{5}{16}   (n_{12}v_1)^4 (n_{12}v_2)^2 \nonumber\\ & & \quad - 
\frac{5}{32} (n_{12}v_1)^3 (n_{12}v_2)^3 +
\frac{19}{16} (n_{12}v_1)^3 (n_{12}v_2) v_1^2 + \frac{15}{16} (n_{12}v_1)^2
(n_{12}v_2)^2 v_1^2 \nonumber\\ & & \quad
+ \frac{3}{4} (n_{12}v_1) (n_{12}v_2)^3 v_1^2 +
\frac{19}{16} (n_{12}v_2)^4 v_1^2 - \frac{21}{16} (n_{12}v_1) (n_{12}v_2)
v_1^4 - 2 (n_{12}v_2)^2 v_1^4 + \frac{55}{16} v_1^6 \nonumber\\ & & \quad
- \frac{19}{16} (n_{12}v_1)^4 (v_1v_2) - (n_{12}v_1)^3 (n_{12}v_2) (v_1v_2) -
\frac{15}{32} (n_{12}v_1)^2 (n_{12}v_2)^2 (v_1v_2) \nonumber\\ & & \quad
+ \frac{45}{16}
(n_{12}v_1)^2 v_1^2 (v_1v_2) + \frac{5}{4} (n_{12}v_1) (n_{12}v_2) v_1^2
(v_1v_2) + \frac{11}{4}(n_{12}v_2)^2 v_1^2 (v_1v_2)  \nonumber\\ & & \quad -
\frac{139}{16} v_1^4 (v_1v_2) - \frac{3}{4} 
(n_{12}v_1)^2 (v_1v_2)^2 + \frac{5}{16} (n_{12}v_1) (n_{12}v_2) (v_1v_2)^2 +
\frac{41}{8} v_1^2 (v_1v_2)^2  \nonumber\\ & & \quad 
+\frac{1}{16} (v_1v_2)^3  - \frac{45}{16}
(n_{12}v_1)^2 v_1^2 v_2^2 - \frac{23}{32} (n_{12}v_1) (n_{12}v_2) v_1^2 v_2^2+
 \frac{79}{16} v_1^4 v_2^2 - \frac{161}{32} v_1^2 (v_1v_2) v_2^2   \bigg) 
\nonumber\\ & & \quad + \frac{G^2 m_1^2 m_2}{r_{12}^2}
\bigg(-\frac{49}{8} (n_{12}v_1)^4 + 
\frac{75}{8} (n_{12}v_1)^3 (n_{12}v_2) - \frac{187}{8} (n_{12}v_1)^2
(n_{12}v_2)^2 \nonumber\\ & & \quad
+ \frac{247}{24} (n_{12}v_1) (n_{12}v_2)^3 + \frac{49}{8}
(n_{12}v_1)^2 v_1^2 + \frac{81}{8} (n_{12}v_1) (n_{12}v_2) v_1^2 -
\frac{21}{4} 
(n_{12}v_2)^2 v_1^2 + \frac{11}{2} v_1^4 \nonumber\\ & & \quad
- \frac{15}{2} (n_{12}v_1)^2
(v_1v_2) - \frac{3}{2} (n_{12}v_1) (n_{12}v_2) (v_1v_2) + \frac{21}{4}
(n_{12}v_2)^2 (v_1v_2) - 27 v_1^2 (v_1v_2) \nonumber\\ & & \quad
+ \frac{55}{2} (v_1v_2)^2 +
\frac{49}{4} (n_{12}v_1)^2 v_2^2 - \frac{27}{2} (n_{12}v_1) (n_{12}v_2) v_2^2
+ \frac{3}{4} (n_{12}v_2)^2 v_2^2 + \frac{55}{4} v_1^2 v_2^2 \nonumber\\ & &
\quad 
- 28 (v_1v_2) v_2^2 + \frac{135}{16} v_2^4\bigg) + \frac{3 G^4 m_1^4 m_2}{8
r_{12}^4}+ \frac{G^4 m_1^3
m_2^2}{r_{12}^4} 
\bigg(\frac{5809}{280} - \frac{11}{3} \lambda - 
\frac{22}{3} \ln \left(\frac{r_{12}}{r'_1} \right)\bigg) \nonumber\\ & & \quad
+ \frac{G^3 m_1^2 m_2^2}{r_{12}^3} \bigg(\frac{547}{12} (n_{12}v_1)^2 - 
\frac{3115}{48} (n_{12}v_1) (n_{12}v_2)  -
\frac{123}{64} (n_{12}v_1)^2 \pi^2  \nonumber\\ & & \quad + \frac{123}{64}
(n_{12}v_1) 
(n_{12}v_2) \pi^2  - \frac{575}{18} v_1^2 +
\frac{41}{64} \pi^2 v_1^2 + \frac{4429}{144} (v_1v_2) - \frac{41}{64} \pi^2
(v_1v_2) \bigg) \nonumber\\ & & \quad +
\frac{G^3  m_1^3  m_2}{r_{12}^3} 
\bigg(-\frac{44627}{840} (n_{12}v_1)^2 + \frac{32027}{840} (n_{12}v_1)
(n_{12}v_2)  
+  \frac{3}{2} (n_{12}v_2)^2 + \frac{24187}{2520} v_1^2 \nonumber\\ & & \quad
- 
\frac{27967}{2520} (v_1v_2) + \frac{5}{4} v_2^2 + 22 (n_{12}v_1)^2 \ln
\left(\frac{r_{12}}{r'_1}  \right) - 22 (n_{12}v_1) (n_{12}v_2) \ln
\left(\frac{r_{12}}{r'_1} \right) \nonumber\\ & &\quad -
\frac{22}{3} v_1^2 \ln \left(\frac{r_{12}}{r'_1} \right) + \frac{22}{3}
(v_1v_2) \ln \left(\frac{r_{12}}{r'_1} \right)\bigg)
\Bigg\}+1\leftrightarrow 2 +{\cal
O}\left(\frac{1}{c^7}\right)\;,\label{25}
\end{eqnarray}
We find that this energy is in agreement with the expression obtained
in Ref. \cite{BFeom} by guess-work starting directly from the
equations of motion. The logarithms $\ln
\big(\frac{r_{12}}{r'_1}\big)$ and $\ln \big(\frac{r_{12}}{r'_2}\big)$
take the form of a gauge transformation of the energy (see Eq. (6.16)
in \cite{BFeom}). Accordingly they will never enter a physical result
such as the circular-orbit energy when expressed in terms of the
orbital frequency of the circular motion (see Ref. \cite{BF00}). Such
is not the case of the constant $\lambda$ which does enter the
invariant energy. The total linear momentum $P^i$ at the 3PN order is
given by

\begin{eqnarray} 
P^i&=& m_1 v_1^i \nonumber \\ &+&  \frac{1}{c^2} \Bigg\{-n_{12}^i \frac{G m_1
m_2}{2 r_{12}} (n_{12}v_1) +  v_1^i \bigg(-\frac{G m_1 m_2}{2 r_{12}} +
\frac{m_1  v_1^2}{2}\bigg) \Bigg\} \nonumber \\ &+& \frac{1}{c^4} \Bigg\{  
n_{12}^i \bigg( \frac{G^2 m_1^2 m_2}{r_{12}^2} \bigg(\frac{29}{4}
(n_{12}v_1) - \frac{9}{4} (n_{12}v_2)\bigg) + \frac{G m_1 m_2}{r_{12}}
\bigg(\frac{3}{8} (n_{12}v_1)^3
\nonumber \\ & & \quad + \frac{3}{8}(n_{12}v_1)^2 (n_{12}v_2) 
-  \frac{9}{8} (n_{12}v_1) v_1^2 - \frac{7}{8} (n_{12}v_2) v_1^2 + \frac{7}{4}
(n_{12}v_1) (v_1v_2) \bigg) \bigg) \nonumber \\ & & \quad 
+ v_1^i \bigg(-\frac{3 G^2 m_1^2
m_2}{r_{12}^2} + \frac{7 G^2 m_1 m_2^2}{2r_{12}^2 } +
\frac{3 m_1 v_1^4}{8} \nonumber \\ & & \quad + \frac{G m_1 m_2}{r_{12}} 
\bigg(\frac{13}{8} (n_{12}v_1)^2 - \frac{1}{4} (n_{12}v_1) (n_{12}v_2) 
\nonumber \\ & & \quad - 
\frac{13}{8} (n_{12}v_2)^2 + 
\frac{5}{8} v_1^2 - \frac{7}{4} (v_1v_2) 
+ \frac{7}{8} v_2^2 \bigg) \bigg)\Bigg\} \nonumber \\ &+& 
\frac{1}{c^6} \Bigg\{ n_{12}^i\bigg( \frac{G^2 m_1^2 m_2}{r_{12}^2}
\bigg(-\frac{45}{8} (n_{12}v_1)^3 
+ \frac{59}{8} (n_{12}v_1)^2 (n_{12}v_2) - \frac{179}{8} (n_{12}v_1)
(n_{12}v_2)^2 \nonumber \\ & & \quad + \frac{247}{24} (n_{12}v_2)^3  +
\frac{135}{8} (n_{12}v_1) v_1^2 
- \frac{87}{8} (n_{12}v_2) v_1^2 - \frac{53}{2} (n_{12}v_1) (v_1v_2) +
\frac{87}{4} (n_{12}v_2) (v_1v_2) \nonumber \\ & & \quad + \frac{53}{4}
(n_{12}v_1) v_2^2  - 12 (n_{12}v_2) v_2^2 \bigg) + \frac{G m_1 m_2}{r_{12}}
\bigg(-\frac{5}{16} (n_{12}v_1)^5 - \frac{5}{16} (n_{12}v_1)^4 (n_{12}v_2) 
\nonumber \\ & &
\quad - \frac{5}{16} (n_{12}v_1)^3 (n_{12}v_2)^2 + \frac{19}{16}
(n_{12}v_1)^3 v_1^2 + \frac{15}{16} (n_{12}v_1)^2 (n_{12}v_2) v_1^2 +
\frac{3}{4} (n_{12}v_1) (n_{12}v_2)^2 v_1^2 \nonumber \\ & & \quad +
\frac{5}{8} (n_{12}v_2)^3 v_1^2 - \frac{21}{16} (n_{12}v_1) v_1^4 -
\frac{11}{16} (n_{12}v_2) v_1^4 - \frac{5}{4} (n_{12}v_1)^3 (v_1v_2) \nonumber
\\ & & \quad - \frac{9}{8} (n_{12}v_1)^2 (n_{12}v_2) (v_1v_2) + \frac{15}{8}
(n_{12}v_1) v_1^2 (v_1v_2) + (n_{12}v_2) v_1^2 (v_1v_2) - \frac{1}{8}
(n_{12}v_1) (v_1v_2)^2 \nonumber \\ & & \quad - \frac{15}{16}
(n_{12}v_1) v_1^2 v_2^2 \bigg) - \frac{175G^3 m_1^2 m_2^2}{8 r_{12}^3}
(n_{12}v_1) + \frac{G^3 m_1^3 m_2}{r_{12}^3} \bigg(-\frac{46517}{840}
(n_{12}v_1) \nonumber \\ & & \quad + \frac{34547}{840} (n_{12}v_2) + 22 
(n_{12}v_1) \ln \left(\frac{r_{12}}{r'_1} \right) - 22 (n_{12}v_2) \ln
\left(\frac{r_{12}}{r'_1} \right) \bigg) \bigg) + v_1^i \bigg(
\frac{5 m_1 
v_1^6}{16} \nonumber \\ & & \quad + \frac{G^2 m_1^2 m_2}{r_{12}^2}
\bigg(-\frac{39}{4} (n_{12}v_1)^2 + 22 (n_{12}v_1) (n_{12}v_2) - \frac{21}{4} 
(n_{12}v_2)^2 - 2 v_1^2 + (v_1v_2) - \frac{1}{2} v_2^2\bigg) \nonumber \\ & &
\quad  + \frac{G^2 m_1 m_2^2}{r_{12}^2} \bigg(\frac{23}{4}
(n_{12}v_1)^2 - \frac{101}{4} (n_{12}v_1) (n_{12}v_2) + 17 (n_{12}v_2)^2 +
\frac{9}{4} v_1^2 - (v_1v_2) + \frac{1}{2}v_2^2 \bigg) \nonumber \\ & & \quad + 
\frac{G m_1 m_2}{r_{12}} \bigg(-\frac{19}{16}
(n_{12}v_1)^4 + \frac{1}{4} (n_{12}v_1)^3 (n_{12}v_2) + \frac{3}{16}
(n_{12}v_1)^2 (n_{12}v_2)^2 \nonumber \\ & & \quad + \frac{1}{8} (n_{12}v_1)
(n_{12}v_2)^3 + \frac{19}{16} (n_{12}v_2)^4 + \frac{45}{16} (n_{12}v_1)^2
v_1^2 - \frac{5}{8} (n_{12}v_1) (n_{12}v_2) v_1^2 
- 2 (n_{12}v_2)^2 v_1^2 \nonumber \\ & & \quad + \frac{23}{16} v_1^4 -
\frac{3}{4} (n_{12}v_1)^2 
(v_1v_2) + \frac{3}{4} (n_{12}v_1) (n_{12}v_2) (v_1v_2) + \frac{19}{8}
(n_{12}v_2)^2 (v_1v_2) - \frac{31}{8} v_1^2 (v_1v_2) \nonumber \\ & & \quad +
\frac{17}{8} (v_1v_2)^2 
+ \frac{3}{8} (n_{12}v_1)^2 v_2^2 - \frac{1}{2} (n_{12}v_1) (n_{12}v_2) v_2^2 -
\frac{45}{16} (n_{12}v_2)^2 v_2^2 + \frac{31}{16} v_1^2 v_2^2 \nonumber \\ & &
\quad - 3 (v_1v_2)
v_2^2 + \frac{19}{16} v_2^4 \bigg) -\frac{19 G^3 m_1^2 m_2^2}{8 r_{12}^3}+
\frac{G^3 m_1^3 m_2}{r_{12}^3} \bigg(\frac{20407}{2520} - \frac{22}{3} \ln
\left(\frac{r_{12}}{r'_1} \right)\bigg) \nonumber \\ & & \quad + 
\frac{G^3 m_1 m_2^3}{r_{12}^3} \bigg(-\frac{21667}{2520} + \frac{22}{3} \ln
\left(\frac{r_{12}}{r'_2} \right)\bigg) \bigg) \Bigg\}+ 1 \leftrightarrow 2+{\cal
O}\left(\frac{1}{c^7}\right)\;, \label{26}
\end{eqnarray}
Next, the 3PN angular momentum $J^i$ is

\begin{eqnarray} 
J^i &=& \varepsilon_{ijk} m_1 y_1^j v_1^k \nonumber \\
&+&\frac{1}{c^2}\varepsilon_{ijk} \Bigg\{ y_1^j v_1^k \bigg(\frac{3 G
m_1 m_2}{r_{12} }+ \frac{m_1 v_1^2}{2}\bigg) - y_1^j v_2^k \frac{7 G
m_1 m_2 }{2 r_{12}} + y_1^j y_2^k \frac{G m_1 m_2}{2 r_{12}^2}
(n_{12}v_1) \Bigg\}
\nonumber \\ &+& \frac{1}{c^4} \varepsilon_{ijk}
\Bigg\{- v_1^j v_2^k \frac{7 G m_1 m_2}{4} (n_{12}v_1) +
y_1^j v_1^k \bigg( -\frac{5 G^2 m_1^2 
m_2}{4 r_{12}^2}+ \frac{7 G^2 m_1 m_2^2}{2r_{12}^2} \nonumber \\ & & \quad +
\frac{3 m_1 v_1^4}{8} + 
\frac{G m_1 m_2}{r_{12}} \bigg(-\frac{3}{2} (n_{12}v_2)^2 + \frac{7}{2} v_1^2 - 4
(v_1v_2) + 2 v_2^2\bigg) \bigg) \nonumber \\ & & \quad +  y_1^j v_2^k  \bigg(-\frac{7 
G^2 m_1 m_2^2}{4 r_{12}^2} + \frac{G m_1 m_2}{r_{12}}
\bigg( -\frac{1}{8} (n_{12}v_1)^2 - \frac{1}{4} (n_{12}v_1) (n_{12}v_2)
\nonumber \\ & & \quad +
\frac{13}{8} (n_{12}v_2)^2 - \frac{9}{8} v_1^2 + \frac{9}{4} (v_1v_2) -
\frac{23}{8} v_2^2 \bigg) \bigg) 
+  y_1^j y_2^k \bigg(\frac{G^2 m_1^2 m_2}{r_{12}^3} \bigg(-\frac{29}{4} 
(n_{12}v_1) \nonumber \\ & & \quad + \frac{9}{4} (n_{12}v_2)\bigg) + \frac{G
m_1 m_2}{r_{12}^2}  \bigg(-\frac{3}{8} (n_{12}v_1)^3 - \frac{3}{8}
(n_{12}v_1)^2 (n_{12}v_2) + \frac{9}{8} (n_{12}v_1) v_1^2 \nonumber \\ & &
\quad + \frac{7}{8}
(n_{12}v_2) v_1^2 - \frac{7}{4} (n_{12}v_1) (v_1v_2) \bigg) \bigg) \Bigg\}
\nonumber \\ &+& \frac{1}{c^6} \varepsilon_{ijk}\Bigg\{ v_1^j v_2^k
\bigg(\frac{G^2 m_1^2 
m_2}{r_{12}} \bigg(\frac{235}{24} (n_{12}v_1) - \frac{235}{24} (n_{12}v_2)
\bigg) + G m_1 m_2 \bigg(\frac{5}{12} (n_{12}v_1)^3 \nonumber \\ & & \quad + 
\frac{3}{8} (n_{12}v_1)^2 (n_{12}v_2) - \frac{15}{8} (n_{12}v_1) v_1^2 -
(n_{12}v_2) v_1^2 + \frac{1}{4} (n_{12}v_1) (v_1v_2) \bigg) \bigg) \nonumber
\\ & & \quad + y_1^j v_1^k \bigg( \frac{5 m_1 v_1^6}{16} +
\frac{G m_1 m_2}{r_{12}} \bigg(\frac{9}{8} (n_{12}v_2)^4 - \frac{7}{4}
(n_{12}v_2)^2 
v_1^2 + \frac{33}{8} v_1^4  \nonumber \\ & & \quad + 2 (n_{12}v_2)^2 (v_1v_2)-
6 v_1^2 (v_1v_2) + 2 
(v_1v_2)^2 - \frac{5}{2} (n_{12}v_2)^2 v_2^2 + 3 v_1^2 v_2^2 - 4 (v_1v_2)
v_2^2 + 2 v_2^4 \bigg) \nonumber \\ & & \quad + \frac{G^2 m_1^2 m_2}{r_{12}^2}
\bigg( -\frac{161}{12} 
(n_{12}v_1)^2 + \frac{223}{12} (n_{12}v_1) (n_{12}v_2) - \frac{29}{12}
(n_{12}v_2)^2 + \frac{41}{24} v_1^2 - \frac{14}{3} (v_1v_2) \nonumber \\ & &
\quad + \frac{7}{3} 
v_2^2\bigg) + \frac{G^2 m_1 m_2^2}{r_{12}^2} \bigg(\frac{1}{2}(n_{12}v_1)^2 -
(n_{12}v_1) (n_{12}v_2) - 3 (n_{12}v_2)^2 + 
\frac{45}{4} v_1^2 - 19 (v_1v_2) \nonumber \\ & & \quad + \frac{19}{2}
v_2^2\bigg) + \frac{5 G^3 m_1 m_2^3}{2 r_{12}^3}+ \frac{G^3 m_1^2
m_2^2}{r_{12}^3} \bigg(-\frac{2605}{72} +\frac{41}{32} \pi^2\bigg) + \frac{G^3 
m_1^3 m_2}{r_{12}^3} \bigg(\frac{55409}{2520} \nonumber \\ & & \quad -
\frac{44}{3} \ln \left(\frac{r_{12}}{r'_1} 
\right) \bigg)  \bigg) + y_1^j y_2^k \bigg( \frac{G^2 m_1^2
m_2}{r_{12}^3} \bigg(\frac{45}{8} (n_{12}v_1)^3 - \frac{59}{8} (n_{12}v_1)^2
(n_{12}v_2) \nonumber \\ & & \quad + \frac{179}{8} (n_{12}v_1) 
(n_{12}v_2)^2 - \frac{247}{24} (n_{12}v_2)^3 - \frac{135}{8} (n_{12}v_1) v_1^2
+ \frac{87}{8} (n_{12}v_2) v_1^2 \nonumber \\ & & \quad + \frac{53}{2}
(n_{12}v_1)  
(v_1v_2) - \frac{87}{4} (n_{12}v_2) (v_1v_2) - \frac{53}{4} (n_{12}v_1) v_2^2
+ 12 (n_{12}v_2) v_2^2\bigg) \nonumber \\ & & \quad + \frac{G m_1
m_2}{r_{12}^2} \bigg(\frac{5}{16}  (n_{12}v_1)^5 + 
\frac{5}{16} (n_{12}v_1)^4 (n_{12}v_2) + \frac{5}{16} (n_{12}v_1)^3
(n_{12}v_2)^2 - \frac{19}{16} (n_{12}v_1)^3 v_1^2 \nonumber \\ & & \quad -
\frac{15}{16}  (n_{12}v_1)^2 (n_{12}v_2) v_1^2 - \frac{3}{4} (n_{12}v_1)
(n_{12}v_2)^2 v_1^2 
- \frac{5}{8} (n_{12}v_2)^3 v_1^2 + \frac{21}{16} (n_{12}v_1) v_1^4 \nonumber
\\ & & \quad +
\frac{11}{16} (n_{12}v_2) v_1^4 + \frac{5}{4} (n_{12}v_1)^3 (v_1v_2) +
\frac{9}{8} (n_{12}v_1)^2 (n_{12}v_2) (v_1v_2) - \frac{15}{8} (n_{12}v_1)
v_1^2 (v_1v_2) \nonumber \\ & & \quad - (n_{12}v_2) v_1^2 (v_1v_2) +
\frac{1}{8} (n_{12}v_1) 
(v_1v_2)^2 + \frac{15}{16} (n_{12}v_1) v_1^2 v_2^2 \bigg) + \frac{175 G^3 m_1^2
m_2^2}{8 r_{12}^4} (n_{12}v_1) \nonumber \\ & & \quad + \frac{G^3 m_1^3
m_2}{r_{12}^4} 
\bigg(\frac{46517}{840} (n_{12}v_1) - \frac{34547}{840} (n_{12}v_2) - 22
(n_{12}v_1) \ln \left(\frac{r_{12}}{r'_1} 
\right) \nonumber \\ & & \quad + 22 (n_{12}v_2) \ln \left(\frac{r_{12}}{r'_1}
\right) \bigg) \bigg) + 
y_1^j v_2^k \bigg( \frac{G^2 m_1^2 m_2}{r_{12}^2} \bigg(20
(n_{12}v_1)^2 - \frac{97}{4} (n_{12}v_1) (n_{12}v_2) \nonumber \\ & & \quad +
\frac{21}{4} (n_{12}v_2)^2 - 9 v_1^2 + 18 (v_1v_2) - 9 v_2^2 \bigg) + 
\frac{G^2 m_1 m_2^2}{r_{12}^2} \bigg( -\frac{17}{6} (n_{12}v_1)^2 \nonumber \\
& & \quad + \frac{41}{12}  (n_{12}v_1) (n_{12}v_2) +
\frac{11}{3} (n_{12}v_2)^2 - \frac{17}{6} v_1^2 + \frac{17}{3} (v_1v_2) -
\frac{89}{24} v_2^2 \bigg) \nonumber \\ & & \quad + \frac{G m_1 m_2}{r_{12}}
\bigg( \frac{1}{16} 
(n_{12}v_1)^4  + \frac{1}{8} (n_{12}v_1)^3 (n_{12}v_2) + \frac{3}{16}
(n_{12}v_1)^2 (n_{12}v_2)^2 + \frac{1}{4} (n_{12}v_1) (n_{12}v_2)^3 \nonumber
\\ & & \quad - \frac{19}{16} (n_{12}v_2)^4 - \frac{5}{16}
(n_{12}v_1)^2 v_1^2 - \frac{1}{2} (n_{12}v_1) (n_{12}v_2) v_1^2 + \frac{3}{8}
(n_{12}v_2)^2 v_1^2 - \frac{13}{16} v_1^4 \nonumber \\ & & \quad + \frac{3}{8}
(n_{12}v_1)^2 (v_1v_2) 
+ \frac{3}{4} (n_{12}v_1) (n_{12}v_2) (v_1v_2) - \frac{3}{4} (n_{12}v_2)^2
(v_1v_2) + v_1^2 (v_1v_2) + \frac{1}{8} (v_1v_2)^2 \nonumber \\ & & \quad -
\frac{1}{4} (n_{12}v_1)^2 
v_2^2 - \frac{5}{8} (n_{12}v_1) (n_{12}v_2) v_2^2 + \frac{45}{16}
(n_{12}v_2)^2 v_2^2 - \frac{17}{16} v_1^2 v_2^2 + \frac{17}{8} (v_1v_2) v_2^2
- \frac{43}{16} v_2^4\bigg) \nonumber \\ & & \quad + \frac{G^3 m_1^2
m_2^2}{r_{12}^3} 
\bigg(\frac{1217}{36}  - \frac{41}{32} \pi^2 \bigg) + \frac{G^3 m_1^3
m_2}{r_{12}^3} \bigg(-\frac{27967}{2520} + \frac{22}{3} \ln
\left(\frac{r_{12}}{r'_1} \right) \bigg) \nonumber \\ & & \quad + \frac{G^3 m_1
m_2^3}{r_{12}^3} 
\bigg(-\frac{17501}{1260} + \frac{22}{3} \ln \left(\frac{r_{12}}{r'_2} 
\right) \bigg) \bigg)\Bigg\} + 1 \leftrightarrow 2 + {\cal O}\left( \frac{1}{c^7}
\right) \;.\label{27}
\end{eqnarray}
The last constant of the motion is the vector $K^i$. We rather present
the vector $G^i=P^it+K^i$ which represents the center-of-mass position
and varies linearly with time.

\begin{eqnarray} 
G^i &=& 
m_1 y_1^i \nonumber \\ &+&  \frac{1}{c^2}  
\Bigg\{y_1^i\bigg(-\frac{G m_1 m_2}{2 r_{12}} + \frac{
m_1 v_1^2}{2}\bigg) \Bigg\} \nonumber \\ &+& \frac{1}{c^4}\Bigg\{ v_1^i G
m_1 m_2 \bigg(-\frac{7}{4} (n_{12}v_1) - 
\frac{7}{4} (n_{12}v_2)\bigg)    +  y_1^i
\bigg(-\frac{5 G^2 m_1^2 m_2}{4 r_{12}^2} + \frac{7 G^2 m_1
m_2^2}{4r_{12}^2} \nonumber  \\ & & \quad +  
\frac{3 m_1 v_1^4}{8}  + \frac{G m_1
m_2}{r_{12}} 
\bigg(-\frac{1}{8} (n_{12}v_1)^2 - \frac{1}{4} (n_{12}v_1) (n_{12}v_2) +
\frac{1}{8} (n_{12}v_2)^2 \nonumber \\ & & \quad + \frac{19}{8} v_1^2  -
\frac{7}{4} 
(v_1v_2) - \frac{7}{8} v_2^2\bigg)\bigg)\Bigg\}    \nonumber \\ &+&
\frac{1}{c^6}  
\Bigg\{ v_1^i \bigg( \frac{235G^2 m_1^2 m_2}{24r_{12}} \bigg(
(n_{12}v_1) - (n_{12}v_2)\bigg) - \frac{235G^2 m_1 m_2^2}{24r_{12}}
 \bigg( (n_{12}v_1) - 
(n_{12}v_2)\bigg) \nonumber \\ & & \quad + G m_1 m_2 \bigg(\frac{5}{12}
(n_{12}v_1)^3 + \frac{3}{8} 
(n_{12}v_1)^2 (n_{12}v_2) + \frac{3}{8} (n_{12}v_1) (n_{12}v_2)^2 
\nonumber \\ & & \quad +
\frac{5}{12} (n_{12}v_2)^3 
- \frac{15}{8} (n_{12}v_1) v_1^2 - (n_{12}v_2)
v_1^2 + \frac{1}{4} (n_{12}v_1) (v_1v_2) \nonumber \\ & & \quad 
+ \frac{1}{4} (n_{12}v_2) (v_1v_2) -
(n_{12}v_1) v_2^2 - \frac{15}{8} (n_{12}v_2) v_2^2\bigg)
\bigg) \nonumber \\ & & \quad +  
y_1^i \bigg( \frac{5m_1 v_1^6}{16} + 
\frac{G m_1 m_2}{r_{12}}
\bigg(\frac{1}{16} (n_{12}v_1)^4 + \frac{1}{8} (n_{12}v_1)^3 (n_{12}v_2) +
\frac{3}{16} (n_{12}v_1)^2 (n_{12}v_2)^2 \nonumber \\ & & \quad  
+ \frac{1}{4}
(n_{12}v_1) (n_{12}v_2)^3 - \frac{1}{16} 
(n_{12}v_2)^4 - \frac{5}{16} (n_{12}v_1)^2 v_1^2 - \frac{1}{2} (n_{12}v_1)
(n_{12}v_2) v_1^2 \nonumber \\ & & \quad - 
\frac{11}{8} (n_{12}v_2)^2 v_1^2 + \frac{53}{16} v_1^4 + \frac{3}{8}
(n_{12}v_1)^2 (v_1v_2) + \frac{3}{4} (n_{12}v_1) (n_{12}v_2) (v_1v_2) +
\frac{5}{4} (n_{12}v_2)^2 (v_1v_2) \nonumber \\ & & \quad - 5 v_1^2 (v_1v_2) +
\frac{17}{8} 
(v_1v_2)^2 - \frac{1}{4} (n_{12}v_1)^2 v_2^2 - \frac{5}{8} (n_{12}v_1)
(n_{12}v_2) v_2^2 + \frac{5}{16} (n_{12}v_2)^2 v_2^2 \nonumber \\ & & \quad +
\frac{31}{16} v_1^2 
v_2^2 - \frac{15}{8} (v_1v_2) v_2^2 - \frac{11}{16} v_2^4\bigg) + \frac{G^2
m_1^2 m_2}{r_{12}^2} \bigg(\frac{79}{12} (n_{12}v_1)^2 - \frac{17}{3}
(n_{12}v_1) (n_{12}v_2) \nonumber \\ & & \quad + 
\frac{17}{6} (n_{12}v_2)^2 - \frac{175}{24} v_1^2 + \frac{40}{3} (v_1v_2) -
\frac{20}{3} v_2^2\bigg) + \frac{G^2 m_1 m_2^2}{r_{12}^2} 
\bigg(-\frac{7}{3} (n_{12}v_1)^2
\nonumber \\ & & \quad +
\frac{29}{12} (n_{12}v_1) (n_{12}v_2) + \frac{2}{3} (n_{12}v_2)^2 +
\frac{101}{12} v_1^2 - 
\frac{40}{3} (v_1v_2) + \frac{139}{24} v_2^2 \bigg) \nonumber 
\\ & & \quad -\frac{19 G^3 m_1^2 m_2^2}{8 r_{12}^3}+ \frac{G^3 m_1^3
m_2}{r_{12}^3} \bigg(\frac{13721}{1260}  - \frac{22}{3} \ln
\left(\frac{r_{12}}{r'_1} \right)\bigg)  \nonumber \\ & & \quad 
+ \frac{G^3 m_1 m_2^3}{r_{12}^3}
\bigg(-\frac{14351}{1260} + 
\frac{22}{3} \ln \left(\frac{r_{12}}{r'_2} \right)\bigg)\bigg) \Bigg\}
+ 1 \leftrightarrow 2 + {\cal O}\left( \frac{1}{c^7} \right)  \;.\label{28}
\end{eqnarray}
We checked that this expression of the harmonic-coordinate center of
mass is changed under the contact transformation into the
ADM-coordinate expression which is given by Eqs. (16)-(22) in
Ref. \cite{DJS00b}. Notice that the energy $E$ is the only one among
these integrals of the 3PN motion that depends on the unknown constant
$\lambda$. The other integrals $P^i$, $J^i$ and $G^i$ do not depend on
$\lambda$ and therefore are entirely determined.

The latter Noetherian quantities are no longer conserved when we take
into account the radiation reaction effect at the 2PN order. In order
to express the resulting balance equations in the best way, we modify
all these quantities by certain terms of order 2.5PN and find that the
right-hand-sides of the equations take the form appropriate to a
radiative flux at infinity. We pose

\begin{mathletters}\label{30}\begin{eqnarray}
{\widetilde E} &=& E+\frac{4 G^2 m_1^2 m_2}{5 c^5 r_{12}^2}  (n_{12}v_1) \bigg[
(v_1-v_2)^2 +\frac{2 G (m_2-m_1)}{r_{12}}
 \bigg] + 1\leftrightarrow 2 \;, \\
{\widetilde P}^i &=& P^i+
\frac{4 G^2 m_1^2 m_2 }{5 c^5r_{12}^2} n_{12}^i\bigg((v_1-v_2)^2 -\frac{2 G
m_1 }{r_{12}} \bigg) + 1 \leftrightarrow 2 \;, \\
{\widetilde J}^i &=& J^i+ \frac{4 G m_1 m_2}{5 c^5}\varepsilon_{ijk} \bigg[ 
 v_1^2 v_1^j v_2^k +2 \frac{G m_1}{r_{12}}  v_1^j
v_2^k -  \frac{2 G
m_1 }{r_{12}^2} (n_{12} v_1)\big( v_1^j y_2^k
+ y_1^j v_1^k \big)  \nonumber \\ & &
\quad   -   \frac{G m_1
}{r_{12}^3}  (v_1-v_2)^2 y_1^j y_2^k+ \frac{2 G^2
m_1^2}{r_{12}^4} y_1^j y_2^k\bigg] + 1 \leftrightarrow 2 \;,  
\\
{\widetilde G}^i &=& G^i+ \frac{4 G m_1 m_2}{5c^5}v_1^i
\bigg( (v_1-v_2)^2 - \frac{2 G (m_1+m_2)}{r_{12}} \bigg)
+ 1 \leftrightarrow 2  \;.
\end{eqnarray}\end{mathletters}$\!\!$
as well as ${\widetilde K}^i= {\widetilde G}^i- t{\widetilde P}^i$.
Then, the 3PN balance equations are given by

\begin{mathletters}\label{29}\begin{eqnarray}
\frac{d\widetilde{E}}{dt} &=& -\frac{G}{5 c^5} \frac{d^3 Q_{ij}}{dt^3}
\frac{d^3 Q_{ij}}{dt^3} + {\cal O}\left( \frac{1}{c^7} \right)\;, \\
\frac{d{\widetilde P}^i}{dt} &=& {\cal O}\left( \frac{1}{c^7} \right)\;,\\
\frac{d{\widetilde J}^i}{dt} &=& - \frac{2G}{5 c^5}\varepsilon_{ijk}
\frac{d^2 Q_{jl} }{dt^2} \frac{d^3 Q_{kl}}{dt^3} 
+ {\cal O}\left( \frac{1}{c^7} \right)\;, \\ 
\frac{d{\widetilde K}^i}{dt} &=& {\cal O}\left( \frac{1}{c^7} \right)\;,
\end{eqnarray}\end{mathletters}$\!\!$
where the Newtonian trace-free quadrupole moment is $Q_{ij}=m_1 (y_1^i
y_1^j-\frac{1}{3}\delta^{ij} y_1^2)+1 \leftrightarrow 2$.

\subsection{Contact transformation and the ADM Hamiltonian at the 3PN order}

Our final result for the contact transformation (\ref{17}) is as
follows. The first term in (\ref{17}) is composed of the conjugate
momentum of the acceleration and is readily obtained by
differentiating (\ref{24}):

\begin{align}\label{31'}
q_1^i &= \frac{1}{c^4}\Bigg\{n_{12}^i\bigg(-\frac{1}{8}G m_1
m_2(n_{12}v_2)^2 +\frac{7}{8}G m_1 m_2 v_2^2\bigg)-\frac{7}{4}G m_1
m_2(n_{12}v_2) v_2^i\Bigg\}\nonumber \\ & +
\frac{1}{c^6} \Bigg\{ n_{12}^i 
\bigg(\frac{G^2 m_1^2 m_2}{r_{12}} 
\bigg( -\frac{17}{6} (n_{12}v_2)^2  + 
\frac{185}{16} v_1^2 -
\frac{185}{8} (v_1v_2)+ 
\frac{20}{3} v_2^2\bigg) \nonumber \\ & \qquad 
 + \frac{G^2 m_1 m_2^2}{r_{12}} \bigg( \frac{29}{24} (n_{12}v_2)^2 +
 \frac{235}{48} v_2^2 \bigg) + G m_1 m_2 \bigg(\frac{1}{8}
 (n_{12}v_1)(n_{12}v_2)^3+\frac{1}{16} (n_{12}v_2)^4\nonumber \\ &
 \qquad +\frac{3}{8} (n_{12}v_2)^2(v_1v_2)-\frac{5}{8}
 (n_{12}v_1)^2v_2^2-\frac{1}{2}
 (n_{12}v_1)(n_{12}v_2)v_2^2-\frac{5}{16} (n_{12}v_2)^2v_2^2\bigg)
 \bigg) \nonumber \\ & \qquad + v_1^i \bigg( G m_1 m_2 \bigg(
\frac{11}{4} (n_{12}v_2) v_1^2 -2(n_{12}v_2) (v_1v_2)+ \frac{15}{8}
(n_{12}v_2) v_2^2\bigg) \bigg)
\nonumber \\ & \qquad +  v_2^i \bigg( - 
\frac{235 G^2 m_1^2 m_2}{24r_{12}} (n_{12}v_2) + \frac{235
G^2 m_1 m_2^2}{24r_{12}} (n_{12}v_2) \nonumber \\ & \qquad
+ G m_1 m_2\bigg(\frac{3}{8} (n_{12}v_1) (n_{12}v_2)^2 
+\frac{5}{12} (n_{12}v_2)^3-(n_{12}v_2) v_1^2\nonumber 
\\ & \qquad +\frac{1}{4} (n_{12}v_2) (v_1v_2)-\frac{15}{8} (n_{12}v_2) 
v_2^2\bigg)\Bigg\} + {\cal O}\left( \frac{1}{c^7} \right)\;.
\end{align}
The second term in (\ref{17}) involves the function $F$ that
constitutes the only possible freedom to adjust in order to match the
harmonic-coordinate and ADM-Hamiltonian formalisms. This $F$ was
uniquely determined as

\begin{align}\label{31}
F &= \frac{1}{c^4} \Bigg\{ \frac{G^2 m_1^2 m_2 }{r_{12}} \bigg( \frac{7}{4}
(n_{12}v_1) - \frac{1}{4} (n_{12}v_2) \bigg) + \frac{G m_1 m_2 }{4} (n_{12}v_2)
v_1^2 \Bigg\} \nonumber \\ & + \frac{1}{c^6} \Bigg\{ \frac{G^2 m_1^2 m_2}{r_{12}}
\bigg( - \frac{91}{144} (n_{12}v_1)^3 + \frac{21}{16} (n_{12}v_1)^2
(n_{12}v_2) - \frac{113}{24} (n_{12}v_1) v_1^2 \nonumber \\ & \qquad +
\frac{35}{8} (n_{12}v_2) v_1^2 + \frac{195}{16} (n_{12}v_1) (v_1v_2) -
\frac{3}{4} (n_{12}v_1) v_2^2 - 
\frac{1}{8} (n_{12}v_2) v_2^2 \bigg) \nonumber \\ & \qquad 
+ G m_1 m_2 \bigg(-\frac{1}{16}
(n_{12}v_1) (n_{12}v_2)^2 v_1^2 - \frac{5}{24}
(n_{12}v_2)^3 v_1^2 \nonumber \\ & \qquad - 
\frac{1}{2} (n_{12}v_2) v_1^4 + \frac{1}{8} (n_{12}v_2) v_1^2 (v_1v_2) +
\frac{5}{16} (n_{12}v_1) v_1^2 v_2^2 \bigg) \nonumber \\ & \qquad + \frac{G^3
m_1^2 m_2^2}{r_{12}^2} 
\bigg( \frac{245}{18} (n_{12}v_1) - \frac{21}{32} (n_{12}v_1) \pi^2 \bigg) 
\nonumber \\ & \qquad + 
\frac{G^3 m_1^3 m_2}{r_{12}^2} \bigg(- \frac{25867}{2520}
(n_{12}v_1) - \frac{3}{4} (n_{12}v_2) + \frac{22}{3}
(n_{12}v_1) \ln 
\left(\frac{r_{12}}{r'_1} \right) \bigg) \Bigg\}\nonumber \\ & \qquad 
+ 1 \leftrightarrow 2+ {\cal O}\left( \frac{1}{c^7} \right)\;.
\end{align}
Notice the dependence of $F$ on the logarithms, {\it viz}

$$\frac{22}{3}\frac{G^3m_1^3m_2}{c^6r_{12}^2}(n_{12}v_1)\ln
\left(\frac{r_{12}}{r'_1}\right)-\frac{22}{3}\frac{G^3m_1m_2^3}
{c^6r_{12}^2}(n_{12}v_2)\ln
\left(\frac{r_{12}}{r'_2}\right)\;,$$
which is necessary in order for the contact transformation to remove
the logarithms of the harmonic-coordinate Lagrangian (\ref{24}). This
result can be checked to be in agreement with the coordinate
transformation given by Eqs. (7.2) in Ref. \cite{BFeom}. The third
term in (\ref{17}) involves a correction term, purely of order 3PN,
which is defined by (\ref{19}). For this term we get

\begin{align}\label{32}
\frac{1}{c^6} X_1^i &= \frac{1}{c^6} \Bigg\{ n_{12}^i 
\bigg( -\frac{G^3 m_1^3 m_2}{r_{12}^2} -
\frac{49}{4} \frac{G^3 m_1^2 m_2^2}{r_{12}^2}
- \frac{3}{4} \frac{G^3 m_1 m_2^3}{r_{12}^2} 
\nonumber \\ & \qquad 
+ \frac{G^2 m_1^2 m_2}{r_{12}} 
\bigg( \frac{11}{8} (n_{12}v_1)^2 - \frac{1}{4}
(n_{12}v_1) (n_{12}v_2) - 
\frac{27}{8} v_1^2 \bigg) \nonumber \\ & \qquad 
 + \frac{G^2 m_1 m_2^2}{r_{12}} \bigg( \frac{3}{8}
(n_{12}v_2)^2 - \frac{1}{8} v_1^2 - \frac{15}{8} 
v_2^2 \bigg) + G m_1 m_2 \bigg(\frac{1}{16} (n_{12}v_2)^2
v_1^2 - \frac{5}{16} 
v_1^2 v_2^2 \bigg) \bigg) \nonumber \\ & \qquad 
+ v_1^i \bigg(  \frac{35G^2 m_1^2 m_2}{8r_{12}}
(n_{12}v_1) + \frac{G^2 m_1 m_2^2}{ r_{12}} \bigg(
-\frac{1}{4} (n_{12}v_1) - 
\frac{3}{2} (n_{12}v_2) \bigg) \nonumber \\ & \qquad 
+ G m_1 m_2 \bigg(
\frac{1}{8} (n_{12}v_1) (n_{12}v_2)^2 - \frac{3}{4}
(n_{12}v_2) v_1^2 +
\frac{7}{4} (n_{12}v_2) (v_1v_2) - \frac{5}{8} (n_{12}v_1) v_2^2 \bigg) \bigg)
\nonumber \\ & \qquad +  v_2^i \bigg( - 
\frac{7 G^2 m_1^2 m_2}{4r_{12}} (n_{12}v_1) + \frac{21
G^2 m_1 m_2^2}{4r_{12}} (n_{12}v_2) + \frac{7 G m_1 m_2}{8} (n_{12}v_2) v_1^2
\bigg)\Bigg\} \nonumber \\ & \qquad + {\cal O}\left( \frac{1}{c^7} \right)\;.
\end{align}
The term $X_2^i$ is obtained by relabeling $1\leftrightarrow 2$. With
those results we obtain the ADM Lagrangian (\ref{21}) which is an
ordinary Lagrangian, not containing any accelerations, and furthermore
not containing any logarithms. Though the investigations in Section
III.B were done with the harmonic-coordinate quantities taken as
``dummy'' variables, we must present here the ADM Lagrangian in terms
of the variables corresponding to the motion in ADM coordinates. We
denote them exactly like in harmonic coordinates but with upper-case
letters, e.g. $R_{12}=|{\bf Y}_1 - {\bf Y}_2|$, ${\bf N}_{12}=({\bf
Y}_1-{\bf Y}_2)/R_{12}$, $(N_{12}V_2)={\bf N}_{12}.{\bf V}_2$.

\begin{align}\label{33}
L^{\rm ADM} &= \frac{G m_1 m_2}{2 R_{12}} + \frac{1}{2} m_1 V_1^2 \nonumber \\
& + \frac{1}{c^2} \Bigg\{- 
\frac{G^2 m_1^2 m_2}{2 R_{12}^2} + \frac{1}{8} m_1 V_1^4 + \frac{G m_1 m_2
}{R_{12}} \bigg( -\frac{1}{4} (N_{12}V_1) (N_{12}V_2) + \frac{3}{2} V_1^2 -
\frac{7}{4} (V_1V_2) \bigg) \Bigg\} \nonumber \\ & + \frac{1}{c^4}
\Bigg\{\frac{G^3 m_1^3 m_2}{4 R_{12}^3}+ \frac{5 G^3 m_1^2 
m_2^2}{8R_{12}^3} + \frac{m_1 V_1^6}{16} + \frac{G^2 m_1^2 m_2}{
R_{12}^2} \bigg( \frac{15}{8}
(N_{12}V_1)^2 + \frac{11}{8} V_1^2 - \frac{15}{4} (V_1V_2) \nonumber \\ &
\qquad + 2 V_2^2 \bigg) +
\frac{G m_1 m_2}{R_{12}} \bigg( \frac{3}{16} (N_{12}V_1)^2 (N_{12}V_2)^2 -
\frac{1}{4} (N_{12}V_1) (N_{12}V_2) V_1^2 - \frac{5}{8} (N_{12}V_2)^2 V_1^2
\nonumber \\ & \qquad + \frac{7}{8} 
V_1^4 + \frac{3}{4} (N_{12}V_1) (N_{12}V_2) (V_1V_2) - \frac{7}{4} V_1^2
(V_1V_2) + \frac{1}{8} (V_1V_2)^2 + \frac{11}{16} V_1^2 V_2^2 \bigg) \Bigg\}
\nonumber \\ & +
\frac{1}{c^6} \Bigg\{-\frac{G^4 m_1^4 m_2}{8 R_{12}^4}+ \frac{G^4 m_1^3
m_2^2}{R_{12}^4}  \bigg(- \frac{993}{140} + \frac{11}{3} 
\lambda + \frac{21}{32} \pi^2\bigg) + \frac{5 m_1 V_1^8}{128}  \nonumber \\ &
\qquad + \frac{G m_1
m_2}{R_{12}} \bigg(- \frac{5}{32} (N_{12}V_1)^3 (N_{12}V_2)^3 + \frac{3}{16}
(N_{12}V_1)^2 
(N_{12}V_2)^2 V_1^2 \nonumber
\\ & \qquad + \frac{9}{16} (N_{12}V_1) (N_{12}V_2)^3 V_1^2 -
\frac{3}{16} (N_{12}V_1) (N_{12}V_2) V_1^4 - \frac{5}{16} (N_{12}V_2)^2 V_1^4
\nonumber
\\ & \qquad + \frac{11}{16} V_1^6 - \frac{15}{32} (N_{12}V_1)^2 (N_{12}V_2)^2 (V_1V_2)
+
\frac{3}{4} (N_{12}V_1) (N_{12}V_2) V_1^2 (V_1V_2) \nonumber
\\ & \qquad - \frac{1}{16}
(N_{12}V_2)^2 V_1^2 (V_1V_2) - \frac{21}{16} V_1^4 (V_1V_2) + \frac{5}{16}
(N_{12}V_1) (N_{12}V_2) (V_1V_2)^2 \nonumber
\\ & \qquad + \frac{1}{8} V_1^2 (V_1V_2)^2 +
\frac{1}{16} (V_1V_2)^3 -
\frac{5}{16} (N_{12}V_1)^2 V_1^2 V_2^2 \nonumber \\ &
\qquad - \frac{9}{32} (N_{12}V_1) (N_{12}V_2)
V_1^2 V_2^2 
+ \frac{7}{8} V_1^4 V_2^2 
- \frac{15}{32} V_1^2 (V_1V_2) V_2^2
\bigg) \nonumber \\ & \qquad 
+ \frac{G^2 m_1^2 m_2}{R_{12}^2} \bigg(-
\frac{5}{12} (N_{12}V_1)^4 - \frac{13}{8} (N_{12}V_1)^3 (N_{12}V_2) - \frac{23}{24} 
(N_{12}V_1)^2 (N_{12}V_2)^2 \nonumber
\\ & \qquad + \frac{13}{16} (N_{12}V_1)^2 V_1^2 + \frac{1}{4}
(N_{12}V_1) (N_{12}V_2) V_1^2 + \frac{5}{6} (N_{12}V_2)^2 V_1^2 +
\frac{21}{16} V_1^4 \nonumber \\ & \qquad - \frac{1}{2} (N_{12}V_1)^2 (V_1V_2)
+ \frac{1}{3} 
(N_{12}V_1) (N_{12}V_2) (V_1V_2) - \frac{97}{16} V_1^2 (V_1V_2) +
\frac{341}{48} (V_1V_2)^2 \nonumber \\ & \qquad 
+ \frac{29}{24} (N_{12}V_1)^2
V_2^2- (N_{12}V_1) 
(N_{12}V_2) V_2^2 + \frac{43}{12} V_1^2 V_2^2 - \frac{71}{8} (V_1V_2) V_2^2 +
\frac{47}{16} V_2^4 \bigg) \nonumber \\ & \qquad + \frac{G^3 m_1^2
m_2^2}{R_{12}^3} \bigg( 
\frac{73}{16} (N_{12}V_1)^2 - 11 (N_{12}V_1) (N_{12}V_2) + \frac{3}{64}
\pi^2 (N_{12}V_1)^2 \nonumber \\ & \qquad  - \frac{3}{64}  \pi^2(N_{12}V_1) (N_{12}V_2) 
-\frac{265}{48} V_1^2 - \frac{1}{64} \pi^2 V_1^2 + \frac{59}{8} (V_1V_2) +
\frac{1}{64} \pi^2 (V_1V_2) \bigg)  \nonumber \\
& \qquad + \frac{G^3 m_1^3 m_2}{R_{12}^3} \bigg(-5
(N_{12}V_1)^2 - \frac{1}{8} 
(N_{12}V_1) (N_{12}V_2) + \frac{173}{48} V_1^2 - \frac{27}{8} (V_1V_2) +
\frac{13}{8} V_2^2 \bigg) \Bigg\}\nonumber \\ & \qquad  + 1 \leftrightarrow 2
+ {\cal O}\left( \frac{1}{c^7} \right)\;.
\end{align}
The corresponding ADM (or, rather, ADM-type \cite{DJS00b})
Hamiltonian is given by the ordinary Legendre transformation
(\ref{23}) as

\begin{align}\label{34}
H^{\rm ADM} &= - \frac{G m_1 m_2}{2 R_{12}}+ \frac{P_1^2}{2 m_1} \nonumber \\
& + \frac{1}{c^2} 
\Bigg\{-\frac{P_1^4}{8 m_1^3} + \frac{G^2 m_1^2 m_2}{2 R_{12}^2} 
+ \frac{G m_1 m_2}{R_{12}} \bigg(\frac{1}{4} \frac{(N_{12}P_1) (N_{12}P_2)}{
m_1 m_2}  - \frac{3}{2}
\frac{P_1^2}{m_1^2} + \frac{7}{4} \frac{(P_1P_2)}{m_1 m_2} \bigg) \Bigg\}
\nonumber \\ & +
\frac{1}{c^4} \Bigg\{ \frac{P_1^6}{16 m_1^5} -\frac{G^3 m_1^3 
m_2}{4 R_{12}^3}- \frac{5 G^3 m_1^2 m_2^2}{8R_{12}^3} + \frac{G^2 m_1^2
m_2}{R_{12}^2} \bigg(-\frac{3}{2} \frac{(N_{12}P_1) (N_{12}P_2)}{m_1 m_2} 
 \nonumber \\ & \qquad +
\frac{19}{4} \frac{P_1^2}{ m_1^2} - \frac{27}{4}\frac{(P_1P_2)}{m_1 m_2}  
+ \frac{5 P_2^2}{2 m_2^2} \bigg) + \frac{G m_1 m_2}{R_{12}} \bigg( 
-\frac{3}{16} \frac{(N_{12}P_1)^2 (N_{12}P_2)^2}{ m_1^2 m_2^2} 
\nonumber \\ & \qquad + 
\frac{5}{8} \frac{(N_{12}P_2)^2 P_1^2}{ m_1^2 m_2^2}  + \frac{5}{8}
\frac{P_1^4}{ m_1^4} - \frac{3}{4}
\frac{(N_{12}P_1) (N_{12}P_2)  (P_1P_2)}{ m_1^2 m_2^2} - 
\frac{1}{8}\frac{(P_1P_2)^2}{ m_1^2 m_2^2}  - \frac{11}{16} \frac{P_1^2
P_2^2 }{ m_1^2 m_2^2} \bigg)\Bigg\} \nonumber \\ & + \frac{1}{c^6}
\Bigg\{-\frac{5 P_1^8}{128 m_1^7} + \bigg( \frac{G^4 m_1^4 m_2}{8 R_{12}^4}+ 
\frac{G^4 m_1^3 m_2^2}{R_{12}^4} \bigg(
\frac{993}{140} - \frac{11}{3} \lambda - \frac{21}{32} \pi^2 \bigg) \bigg)
\nonumber \\ & \qquad +
\frac{G^3 m_1^2 m_2^2}{R_{12}^3} \bigg( -\frac{43}{16} \frac{(N_{12}P_1)^2}{
m_1^2}  + \frac{119}{16} \frac{(N_{12}P_1) (N_{12}P_2)}{m_1 m_2} 
- \frac{3}{64} \pi^2 \frac{(N_{12}P_1)^2}{ m_1^2} \nonumber \\ & \qquad +
\frac{3}{64}  \pi^2 \frac{(N_{12}P_1) (N_{12}P_2)}{m_1 m_2} - \frac{473}{48}
\frac{P_1^2}{m_1^2} + 
\frac{1}{64} \pi^2 \frac{P_1^2}{m_1^2} + \frac{143}{16} \frac{(P_1P_2)}{m_1
m_2}  - \frac{1}{64}  \pi^2 \frac{(P_1P_2)}{m_1 m_2} \bigg) \nonumber \\ &
\qquad + \frac{G^3 m_1^3 m_2}{R_{12}^3}\bigg( \frac{5}{4}
\frac{(N_{12}P_1)^2}{ m_1^2} + \frac{21}{8} \frac{(N_{12}P_1) (N_{12}P_2)}{
m_1 m_2} - \frac{425}{48} \frac{P_1^2}{m_1^2} + \frac{77}{8} \frac{(P_1P_2) 
}{ m_1 m_2}- \frac{25 P_2^2}{8 m_2^2} \bigg) \nonumber \\ & \qquad + \frac{G^2
m_1^2 m_2}{R_{12}^2}  
\bigg( \frac{5}{12} \frac{(N_{12}P_1)^4}{ m_1^4} - \frac{3}{2}
\frac{(N_{12}P_1)^3 (N_{12}P_2)}{ m_1^3 m_2} + \frac{10}{3} \frac{(N_{12}P_1)^2
(N_{12}P_2)^2}{ m_1^2 m_2^2} \nonumber \\ & \qquad + \frac{17}{16}
\frac{(N_{12}P_1)^2 P_1^2}{ 
m_1^4}  - \frac{15}{8} \frac{(N_{12}P_1) (N_{12}P_2) P_1^2}{ m_1^3 
m_2}  - \frac{55}{12} \frac{(N_{12}P_2)^2 P_1^2}{ m_1^2 m_2^2} +
\frac{P_1^4}{16 m_1^4} \nonumber \\ & \qquad - \frac{11}{8}
\frac{(N_{12}P_1)^2 (P_1P_2)}{ m_1^3 
m_2}   +  \frac{125}{12} \frac{(N_{12}P_1) (N_{12}P_2) (P_1P_2)}{ m_1^2 m_2^2}
- \frac{115}{16}  \frac{P_1^2 (P_1P_2) }{m_1^3 m_2} \nonumber \\ & \qquad 
+ \frac{25}{48}
\frac{(P_1P_2)^2}{ m_1^2 m_2^2} - \frac{193}{48}
\frac{(N_{12}P_1)^2 P_2^2 }{m_1^2 m_2^2} + \frac{371}{48} \frac{P_1^2 P_2^2}{
m_1^2 m_2^2} - \frac{27}{16} \frac{ 
P_2^4}{ m_2^4} \bigg) \nonumber \\ & \qquad 
+ \frac{G m_1 m_2}{R_{12}}  \bigg( \frac{5}{32}
\frac{(N_{12}P_1)^3 (N_{12}P_2)^3}{m_1^3 
m_2^3} + \frac{3}{16} \frac{(N_{12}P_1)^2 (N_{12}P_2)^2
P_1^2}{ m_1^4 m_2^2} \nonumber \\ & \qquad 
-  \frac{9}{16} \frac{(N_{12}P_1) (N_{12}P_2)^3 P_1^2}{ m_1^3 m_2^3}
-\frac{5}{16}  \frac{(N_{12}P_2)^2 P_1^4}{ m_1^4 m_2^2} - \frac{7}{16}
\frac{P_1^6}{m_1^6} \nonumber \\ & \qquad  + \frac{15}{32} \frac{(N_{12}P_1)^2
(N_{12}P_2)^2 (P_1P_2)}{ m_1^3  m_2^3} + \frac{3}{4}
\frac{(N_{12}P_1) (N_{12}P_2) P_1^2 (P_1P_2)}{m_1^4 m_2^2} \nonumber \\ & \qquad  
+ \frac{1}{16}
\frac{(N_{12}P_2)^2 P_1^2 (P_1P_2)}{ m_1^3 m_2^3}-
\frac{5}{16} \frac{(N_{12}P_1) (N_{12}P_2) (P_1P_2)^2}{ m_1^3 m_2^3}
\nonumber \\ & \qquad   +
\frac{1}{8} \frac{P_1^2 (P_1P_2)^2}{m_1^4 m_2^2} 
- \frac{1}{16} \frac{(P_1P_2)^3}{ m_1^3 m_2^3}  
-\frac{5}{16}  \frac{(N_{12}P_1)^2 P_1^2 P_2^2}{m_1^4
m_2^2} \nonumber
\\ & \qquad + \frac{7}{32} \frac{(N_{12}P_1) (N_{12}P_2) P_1^2 P_2^2}{m_1^3 m_2^3}
+ \frac{1}{2} \frac{P_1^4 P_2^2}{ m_1^4 m_2^2}   
 + \frac{1}{32} \frac{P_1^2
(P_1P_2) P_2^2}{  m_1^3 m_2^3} \bigg) \Bigg\}\nonumber \\ & \qquad
+ 1 \leftrightarrow 2 + {\cal O}\left( \frac{1}{c^7} \right)\;.
\end{align}
This result is in perfect agreement with the expression obtained by
Damour, Jaranowski and Sch\"afer \cite{DJS00b}. (Note that in their
published result, Eq. (12) in Ref. \cite{DJS00b}, the following terms
are missing:

$$\frac{G^2}{c^6r_{12}^2}\bigg(-\frac{55}{12}m_1-\frac{193}{48}m_2\bigg)
\frac{(N_{12}P_2)^2
P_1^2}{m_1m_2}+1\leftrightarrow 2\;.$$ This is a misprint which has
been corrected in an Erratum \cite{DJS00b}.)  Finally, we recall that
the agreement works if and only if our undetermined constant $\lambda$
is related to their static-ambiguity constant $\omega_{\rm static}$ by
Eq. (\ref{1}), and their kinetic-ambiguity constant takes the value
$\omega_{\rm kinetic}=\frac{41}{24}$. This completes the proof of the
equivalence of the harmonic-coordinate and ADM-Hamiltonian approaches
to the equations of motion of compact binaries at the 3PN order.

\acknowledgements
We thank Bala R. Iyer for discussions. V.C. de Andrade would like to
thank FAPESP-BRAZIL for financial support. G. Faye acknowledges
financial support of the EU-Network HPRN-CT-2000-00137. The algebraic
computations reported in this paper were performed with the help of
the software Mathematica/Mathtensor.

\end{document}